\documentclass[12pt]{iopart}
\usepackage{graphicx}
\usepackage{iopams}  
\usepackage{hyperref}
\usepackage{enumitem}
\usepackage{todonotes}
\setlist[enumerate]{label={\arabic*.}}

\begin{document}

\title[BNS Merger Simulations Insights]{Insights into Binary Neutron Star Merger Simulations: A Multi-Code Comparison}


\author{Maria C. Babiuc Hamilton$^1$ and William A. Messman$^2$\footnote[1]{Undergraduate student}}

\address{$^1$Department of Mathematics and Physics, Marshall University, Huntington, WV, 25755, USA}
\address{$^2$Department of Physics and Astronomy, Purdue University, West Lafayette, IN, 47907, USA}
\ead{$^1$babiuc@marshall.edu; $^2$wmessman@purdue.edu}
\vspace{10pt}

\begin{abstract}
Gravitational Wave (GW) signals from Binary Neutron Star (BNS) mergers provide critical insights into the properties of matter under extreme conditions. Due to the scarcity of observational data, Numerical Relativity (NR) simulations are indispensable for exploring these phenomena, without replacing the need for observational confirmation. However, simulating BNS mergers is a formidable challenge, and ensuring the consistency, reliability or convergence, especially in the post-merger, remains a work in progress. 
In this paper we assess the performance of current BNS merger simulations by analyzing open-source GW waveforms from five leading NR codes -- \texttt{SACRA}, \texttt{BAM}, \texttt{THC}, \texttt{Whisky}, and \texttt{SpEC}. We focus on the accuracy of these simulations and on the effect of the equation of state (EOS) on waveform predictions. We first check if different codes give similar results for similar initial data, then apply two methods to calculate convergence and quantify discretization errors. Lastly, we perform a thorough investigation into the effect of tidal interactions on key frequencies in the GW spectrum. 
We introduce a novel quasi-universal relation for the transient post-merger time, enhancing our understanding of remnant dynamics in this region. 
This detailed analysis clarifies agreements and discrepancies between these leading NR codes, and highlights necessary improvements  for the advanced accuracy requirements of future GW detectors.
\end{abstract}

%
\vspace{2pc}

\section{\label{sec:intro}Introduction}

When two neutron stars collide, they release gravitational waves and electromagnetic signals across the entire spectrum, from radio to gamma rays.   
This treasure trove of information offers insights into the state of matter under extreme conditions, the engine powering gamma-ray bursts, and the formation of elements beyond iron through the rapid nucleosynthesis process.
The landmark detection of Gravitational Waves (GWs) from the Binary Neutron Star (BNS) merger GW170817 \cite{arXiv:1710.05832} was accompanied by the gamma-ray burst GRB 170817A \cite{arXiv:1710.05834} and ignited the bright kilonova explosion AT2017gfo, fueled by nuclear reactions \cite{arXiv:1710.05443}, which dispersed large quantities of heavy metals throughout the universe. 
The detection of the GW190425 event \cite{arXiv:2001.01761}, from an unusually massive BNS merger, along with several neutron star-black hole merger observations \cite{arXiv:2404.04248, arXiv:2106.15163}, indicate that the population of neutron stars is more varied than previously assumed \cite{arXiv:1603.02698}.
Unfortunately, GW detections from BNS merger systems remain rare. 
Given the scarcity of such observational data, Numerical Relativity (NR) simulations are indispensable for explorations into key aspects of neutron star collisions, while observational confirmation remains essential.

NR codes evolve Einstein's equations of General Relativity (GR) along with Hydro-Dynamics (HD) equations, Maxwell's equations of electromagnetism, and various models for the still unknown nuclear Equation of State (EOS) describing the state of matter at super-nuclear densities in the interior of neutron stars. These simulations are computationally demanding and complex, requiring interdisciplinary expertise across multi-physics, mathematics, and computer science, as noted in several comprehensive reviews \cite{arXiv:1611.01519, arXiv:1808.06011, arXiv:2002.03863, arXiv:2004.02527}. 
Accurate numerical data is necessary for extracting the correct physics from current and future GW signals detections, and for predicting and identifying the possible electromagnetic counterparts \cite{arXiv:2108.07277, arXiv:2307.05376}.

As detection capabilities expand with future terrestrial \cite{arXiv:2307.05376, arXiv:2307.10421, arXiv:2402.05056} and space-based GW detectors  \cite{arXiv:2307.16628, arXiv:2311.01300}, the demands on the accuracy of NR simulations increase significantly \cite{arXiv:1902.09485, arXiv:2110.06957, arXiv:2203.00623, arXiv:2210.09259, arXiv:2401.15331}.
The next generation of GW detectors are expected to discover thousands of BNS mergers, and to offer access into the post-merger emission, enabling us to decipher the still unknown EOS \cite{arXiv:2004.11334, arXiv:2301.09672, arXiv:2310.06025}. 
This will demand new standards for the precision and computational power of NR simulations that surpass the current capabilities \cite{arXiv:2203.08139}.
A close comparison among the results from independent codes that simulate BNS mergers is necessary to better understand their performance and to prepare current codes to meet the sensitivity required to probe new physics \cite{arXiv:2108.12368}.

Comparing results from different numerical simulations requires careful consideration of multiple sources of uncertainty and numerical artifacts. Differences in the initial conditions, physical assumptions and numerical schemes, can all contribute to discrepancies in the predicted waveform. 
By addressing these challenges, we can better understand the extent to which observed differences in the GW data from numerical simulations are due to physical effects rather than numerical discrepancies.
These insights help guide ongoing efforts to improve initial conditions, refine numerical algorithms, reduce truncation errors, and mitigate residual eccentricity, thus advancing our understanding of BNS merger physics.

The objective of this paper is to provide the community with a comprehensive assessment of the current state in BNS mergers simulations. We collect and analyze open-source GW waveforms from five leading NR codes to evaluate their reliability and the impact of the EOS on their waveform predictions. Initially, we verify if similar initial data yield comparable results across different codes. Then, we apply specialized techniques, originally designed to assess errors in computational fluid dynamics, to calculate code convergence and  quantify discretization errors within these codes. After addressing this challenging task, we systematically extract and analyze key frequencies of the GW spectrum, both in the strain and Weyl curvature scalar. This enables us to rigorously investigate the dependence of these frequencies on the tidal interactions between neutron stars, and to quantify the ability of these codes to predict the EOS. We identify a new relation between the EOS and the transient post-merger time, providing deeper insight into the capability of codes to resolve the early post-merger dynamics.
 
Our paper is organized as follows. We begin by providing in Section \ref{sec:review} a background on the field, highlighting the significant advancements in simulating BNS systems and the ongoing challenges, particularly the need for thorough comparative studies between different NR codes. 
Following this, we introduce in Section \ref{sec:QUR} the concept of tidal deformability in BNS mergers and its influence on the waveform, specifically through Quasi-Universal Relations (QUR) that link tidal deformability to certain characteristic frequencies in the GW spectrum.
Next, we review in Section \ref{sec:numerics} the complex numerical techniques commonly implemented in NR codes for simulating BNS mergers. We then briefly summarize in Section \ref{sec:codes} the five codes used in our study and describe the datasets they provide. 
In Section \ref{sec:methods}, we outline our methods, including a new technique for calculating the convergence of BNS codes that handles shocks, discontinuities, and other numerical artifacts. We also explain our approach for comparing different codes, focusing on their accuracy in reproducing known QURs and we introduce a new QUR for time, relevant for describing the transient phase immediately after the merger.
Following this, in Section \ref{sec:results} we report our results. We evaluate code performance based on three criteria: consistency in producing comparable results from similar initial data, success in achieving convergence, and accuracy in predicting tidal effects.
Finally, in Section \ref{sec:conclusion}, we summarize our findings and discuss their implications for future code development.

In this paper, we denote the total mass of the BNS system as $M = M_1 + M_2$, where $M_{1}$ and $M_{2}$ are the individual masses expressed in units of the solar mass $M_{\odot}$. The mass ratio is given by $q = M_2/M_1 \le 1$ and symmetric mass ratio $\eta = M_1M_2/M^{2}$. 
We adopt geometrical or code units, where $G=c= 1$, with $G$ representing the gravitational constant, and $c$ the speed of light.
Physical units can be recovered using the following transformations: $u_M = 1.98841\times 10^{30}~\texttt{kg}$ for the mass, $u_L = u_M G/c^2 = 1.476625038~\texttt{km}$ for distance, and $u_t = u_L/c = 4.925490947 \times 10^{-6}~\texttt{s}$ for time.

We acknowledge that the values of $G$ and $M_{\odot}$ are not known with absolute precision, leading to a potential $0.01\%$ difference across results reported by various codes, depending on the accuracy used for these constants \cite{arXiv:2404.11346}.

\section{\label{sec:review} Advancements and Challenges}

Since the pioneering simulations of BNS mergers \cite{arXiv:gr-qc/9911058}, the field has seen remarkable advancements. 
Follow-up studies showed that the merger outcomes and final states are greatly influenced by the compactness and thus the EOS of the neutron stars \cite{arXiv:gr-qc/0203037} and the strength of the magnetic field \cite{arXiv:1009.2468}, setting the stage for future research. 
Advanced numerical techniques were developed to simulate multidimensional relativistic hydrodynamics \cite{arXiv:gr-qc/0003101}, addressing challenges such as capturing shocks \cite{arXiv:astro-ph/0210618} and handling oscillations \cite{arXiv:1604.07999}. 
Progress in conservative General-Relativistic Magneto-Hydro-Dynamics (GRMHD) solvers \cite{arXiv:astro-ph/0512420} and neutrino cooling mechanisms \cite{arXiv:1105.2125, arXiv:1306.4953}, further enhanced the accuracy of the simulations.
Successive improvements in GRMHD simulations explored scenarios with realistic EOS, mass ratios and magnetic fields \cite{arXiv:1604.03445, arXiv:1607.01791}. 
This enabled more detailed studies exploring cases with long-lived remnants of BNS mergers \cite{arXiv:1701.08738}, the r-process nucleosynthesis of heavy elements \cite{arXiv:1703.06216}, and methods to increase the accuracy of the calculated waveforms \cite{arXiv:1708.08926}.

The GW detection from the GW170817 event drove the field forward, prompting detailed investigations into the fate of the remnant \cite{arXiv:1710.01311, arXiv:1711.02093}, the specifics of the EOS \cite{arXiv:1710.07579, arXiv:1807.06437}, and the electromagnetic characteristics of the jet and its afterglow \cite{arXiv:1808.04831, arXiv:1905.02665}.
The progress continued with improved numerical methods \cite{arXiv:1804.02003} and robust schemes for relativistic ideal MHD, enabling the quantifications of errors that lead to unphysical states \cite{arXiv:2005.01821, arXiv:2202.08839}.
Studies expanded to simulations exploring prompt accretion-induced collapse in asymmetric mergers \cite{arXiv:2003.06015}, and the link between the threshold mass above which the remnant collapses and the maximum mass supported by a given EOS \cite{arXiv:2008.04333}.

Progress continued with the development of handoff tools for transitioning between numerical grids, enhancing the  long-term post-merger simulations \cite{arXiv:2112.09817}. 
New EOS parameterizations, including quark phase transitions \cite{arXiv:1908.03135} and enthalpy-based models \cite{arXiv:2301.13818}, were introduced to better capture the properties of nuclear matter inside neutron stars.
The vast array of simulations fueled the development of analytical GW models for inspiraling BNS, both in the frequency \cite{arXiv:1802.06518} and time domains \cite{arXiv:1806.01772}, extended with NR-informed post-merger models \cite{arXiv:1908.11418, arXiv:2205.09112}. 
Further studies extended the parameter space with simulations of high-mass BNS mergers leading to the rapid formation of low-mass stellar black holes \cite{arXiv:2301.09635, arXiv:2402.11013}.

The numerical simulations of BNS mergers also uncovered empirical relations connecting certain quantities depending on the internal structure of the neutron stars with characteristic parameters describing the emitted GW \cite{arXiv:1106.1616, arXiv:1306.4065}. 
These relations were named Quasi-Universal Relations (QUR), because they appeared to hold consistently across a wide range of theoretical models chosen for the EOS, thus enabling the extraction of neutron stars physical properties from the GW signals without detailed knowledge of their internal composition.
For example, studies showed that once the post-merger GW signal is detected with accuracy by advanced detectors, measurements of its dominant post-merger frequencies can help inform about the neutron star radius \cite{arXiv:1705.10850}, constrain the maximum neutron star mass \cite{arXiv:1711.00040, arXiv:1711.00314}, and can offer a way to extract the EOS \cite{arXiv:1804.03221, arXiv:1907.02424}. 
Motivated by this relevance of QUR, over the last years the NR community has made considerable progress in their exploration \cite{arXiv:1907.08534, arXiv:1912.01461}.

However, systematic analysis of large sets of NR simulations revealed significant variations in the universality of these relations, depending on factors like spin, mass ratio, total mass, and resolution \cite{arXiv:1402.6244, arXiv:1604.00246, arXiv:1708.08926, arXiv:1907.03790, arXiv:2210.16366, arXiv:1806.01625}. 
It was reported that errors can accumulate in longer simulations, making QUR dependent on the initial separation of the binary \cite{arXiv:1605.03424}, and that mass ratio and increased total mass can deteriorate QUR and may even break their universality \cite{arXiv:1608.06187}. 
Moreover, various other physical factors such as magnetic fields, neutrino radiation, and thermal effects can affect the reliability of QUR, leading to significant systematic deviations in post-merger \cite{arXiv:2102.12455, arXiv:2103.02365, arXiv:2106.08023, arXiv:2201.03594, arXiv:2209.02757}. However, the quantitative differences in the QUR introduced by finite-temperature and neutrino effects are typically subdominant compared to finite-resolution uncertainties \cite{arXiv:2111.14858}.

Current research is focused on refining the study of QURs to more accurately quantify their degree of universality
\cite{arXiv:2012.12151}, assess their effectiveness in extracting information on the EOS from the GW signal \cite{arXiv:2310.10854}, probe for new and more accurate relations \cite{arXiv:2310.10728}, and apply them to different EOS models  \cite{arXiv:2402.10868}.
The effectiveness of relying on QUR for parameter extraction from data expected from the next-generation detectors was critically examined in \cite{arXiv:2402.01948} and was found that this method is sub-optimal for constraining the EOS.
This highlights the need to test the ability of the codes to reproduce current QUR and to expand to new relations, ensuring they can adapt to the improved sensitivity and data quality of future detectors.

The trajectory of BNS merger simulations demonstrates a remarkable evolution from foundational studies to complex multi-physics models with increased accuracy and realism. 
Despite these impressive advancements, challenges such as achieving numerical convergence, especially in the post-merger phase persist, motivating ongoing efforts to improve the accuracy and convergence of  simulations \cite{arXiv:1907.03790, arXiv:2204.00698}.

Although there is a critical need for systematic comparative studies between codes, these are notably scarce. Initial comparisons between two independent codes \cite{arXiv:1007.1754} indicated variations of up to $10\%$, underscoring the necessity to refine their accuracy.
This was followed by a collaborative effort to thoroughly analyze matter effects on the waveform from BNS mergers \cite{arXiv:1306.4065}. 
Another study examined the impact of the choice of initial data solver on the error budget in the GW data,  revealing that the GW phases differed by $0.5\texttt{rad}$ after only three orbits, despite an initial difference of only $0.02\%$ \cite{arXiv:1605.07205}.
The phase error associated with two different analytical prescriptions of the EOS was analyzed in \cite{arXiv:1908.05277}.
The group that released the Computational Relativity (CoRe) database \cite{arXiv:1806.01625, arXiv:2210.16366} -- the largest open-source BNS simulations to date -- also performed a comparison between three codes for a few chosen datasets and reported consistent results, finding that the differences between the codes were below estimated uncertainties within each code. 
A targeted comparison of two analytical BNS waveform models and two independent NR codes was conducted in \cite{arXiv:1806.01772}, showing consistent results and good agreement within the estimated error bars.
However, a comprehensive analysis focusing on the differences and uncertainties within five open-source codes \cite{arXiv:2210.13481}, reported lower than the expected $2^{nd}$ order convergence for tests with realistic EOS, a general lack of quantitative agreement in the merger time of $\approx 13.6\%$, and in the post-merger GW signal up to $20\%$. 
The largest discrepancy was found in the waveform immediately after merger, in the survival time of the Hyper-Massive Neutron Star (HMNS) remnant and in the amount of ejecta.
It was also reported that higher-order numerical schemes for GR hydrodynamics lead to convergent waveforms; however, none of the schemes tested so far has achieved the formal high-order accuracy expected for smooth flow \cite{arXiv:2202.08839}.

Although these comparisons provided valuable insights into the reliability and performance of several NR codes simulating BNS mergers, more extensive assessments are needed to improve the fidelity of the GW models, especially in the post-merger. 
Thorough and methodical comparisons of simulated GW data are vital to establish the reliability of the codes and validate their results. 
Performing a meaningful comparison of NR codes that simulate BNS mergers presents a formidable challenge, due to the multitude of factors that can influence the accuracy and performance of the simulations.

In light of these challenges, we conduct a comprehensive evaluation of the GW data produced by five major NR codes that simulate BNS mergers. Our study aims to rigorously assess the accuracy of these simulations, validate their results, and quantify both numerical errors and discrepancies observed between different codes. We perform an extensive quantitative comparison, focusing on the ability of the codes to achieve consistent convergence, and the accuracy of the waveforms in reproducing QUR between GW parameters -- such as merger amplitude, key frequencies and transient time in the post-merger -- and the EOS model.
The importance of code convergence and the role of QURs in extracting EOS-related information from BNS mergers further motivates this study. By validating the codes, we aim to improve the reliability of GW data analysis in constraining the EOS of neutron stars.
It is worth noting that we do not re-evaluate the five codes previously tested by \cite{arXiv:2210.13481}, but instead focus on a different set of codes. This approach broadens the scope of comparative analysis, offering fresh insights into the performance and accuracy of leading numerical relativity codes.

\section{\label{sec:QUR} Quasi-Universal Relations}

GWs from the early BNS inspiral are indistinguishable from the signal sourced by a binary black hole system, but in the late inspiral tidal effects become important and are reflected as a phase shift in the emitted waveform \cite{Damour:1984, arXiv:0711.2420, arXiv:0906.1366, arXiv:1106.1616}. 
As they emit GWs and approach each other, the stars in the binary interact through tidal forces that induce a mass polarization. This  causes a deviation from the spherical symmetry of the stars, which gives rise to a tidal quadrupole moment: 
\begin{equation}
Q_{ij} = -k_2C_{ij}.
\end{equation}
Here, $C_{ij}$ is the tidal tensor of the gravitational potential responsible for the tidal deformation, and $k_2$ is the tidal coupling constant, or the quadrupole Love number \cite{arXiv:0709.1915}, describing the degree with which a star suffers a deformation due to the influence of the companion star.
This tidally-induced quadrupole moment adds to the dominant orbital quadrupole moment. 
GWs are generated by the evolution in time of the total quadrupole moment, and as result, the tidal effects will be imprinted in the emitted GWs, causing a departure of the waveform from the binary-black hole signal \cite{arXiv:1101.1673}.

A related quantity, the tidal deformability $\Lambda$, rescales the Love number with the compactness of the star:
\begin{equation}
\Lambda_i = \frac{2}{3}k_2 \left (\frac{1}{C_i} \right)^5 ;~~i = 1, 2.
\end{equation}
The compactness of star $i$, defined as $C_i = M_{i}/R_{i}$,  with mass $M_{i}$ and radius $R_{i}$, depends on the pressure gradient inside the star, thus on the EOS. 
The overall or effective tidal deformability of the binary system is the mass-weighted average of the individual tidal deformabilities:
\begin{equation}
\tilde \Lambda = \frac{16}{13}\sum_{i=1,2} \Lambda_i \frac{M_i^4}{M^4}\left (12 -11  \frac{M_i}{M} \right ).
\end{equation}

The influence of the chosen EOS in shaping the GW signal is contained in this effective tidal deformability $\tilde \Lambda$, with a larger value corresponding to a stiffer EOS and a larger radius, while a smaller value indicates a softer EOS and a smaller radius. 
The correlations between the effective tidal deformability and certain characteristic frequencies in the GW spectrum are encapsulated in the QURs found empirically by numerical simulations of BNS mergers, similar to the universal relations found for individual neutron stars \cite{arXiv:1302.4499, arXiv:1303.1528, arXiv:1309.3885}.
For example, \cite{arXiv:1402.6244} revealed the existence of a QUR at the last stable orbit of a BNS system, attributed to the conservative dynamics before the merger, with the expectation that as the interaction becomes tidally dominant, these relations will change in a nontrivial fashion. 
Another QUR, first reported in \cite{arXiv:1306.4065}, captures the strong correlation of the GW frequency at the merger $f_{mrg}$, defined at the peak of the GW strain amplitude, with the effective tidal deformability $\tilde \Lambda$.
This relation  between $\tilde \Lambda$ and $f_{mrg}$ was linked to the well known set of universal relations previously identified in \cite{arXiv:1302.4499}, and was further analyzed in \cite{arXiv:1402.6244, arXiv:1412.3240, arXiv:1504.01764, arXiv:1604.00246, arXiv:1907.03790}.
Studies showed that right after the merger, the GW frequency keeps increasing to a local maximum $f_{max}$ around the point where the amplitude reaches a minimum, but not all codes succeed in modeling this transient feature, because unphysical artifacts due to resolution limitations and numerical errors makes this QUR less reliable \cite{arXiv:1306.4065, arXiv:1504.01764, arXiv:1707.03368}. 

An important QUR, first pointed out in \cite{arXiv:1106.1616}, is contained in the dependence between $\tilde \Lambda$ and the main frequency reached by the GW in the post-merger, denoted by the symbol $f_{2}$. 
This frequency was identified as the fundamental quadrupolar $(l=m=2)$ oscillation mode of the fluid after merger, or the f-mode frequency that leaves an imprint in the GW signal if the merger remnant does not undergo prompt collapse \cite{arXiv:1105.0368}.
In this case, after the initial short transient in the post-merger, the remnant enters a quasi-stationary phase and the GW signal becomes approximately monochromatic, with the main power emitted in this f-mode frequency \cite{arXiv:1502.03176, arXiv:1604.00246}.
The relationship between the $f_2$ frequency and the effective tidal deformability $\tilde \Lambda$ was found to hold with different degrees of reliability \cite{arXiv:1408.3789, arXiv:1509.08804, arXiv:1607.06636, arXiv:1707.03368, arXiv:1901.03779}, although the spins of the neutron stars seem to affect these relations \cite{arXiv:1906.05288}.
Follow-up studies continued to explore the connection between the EOS and other characteristic frequencies in the post-merger spectra, identifying an additional $f_1$ frequency, as well as a third and fourth frequency mode \cite{arXiv:1412.3240, arXiv:1502.03176}. 
These modes were attributed to the nonlinear oscillations of the two cores as they collide and bounce repeatedly during the BNS merger \cite{arXiv:1105.0368}.
It was shown however that the $f_1$ frequency is a sideband of $f_2$, appearing as a modulation, and can be found only for nearly symmetric binaries, being more difficult to identify as the symmetric mass ratio decreases \cite{arXiv:1907.03790}. 

The dependence of the \emph{key} frequencies $f_{mrg}$, $f_{max}$ and $f_{2}$ in the GW merger and post-merger spectra with the effective tidal deformability $\tilde \Lambda$ is expressed in general with the empirically fitted formula: 
\begin{equation}
\log_{10}(f_{key}/\texttt{kHz}) = f(\tilde \Lambda^{1/5}, M).
\label{eq:QUR}
\end{equation}

It is well known that the fate of the merger remnant depends on both the EOS and the total mass of the BNS system. 
For example, increased stiffness of the EOS causes faster merger, but longer post-merger duration, and diminishes the value of $f_2$ \cite{arXiv:1605.03424}, while a phase transition after the merger might soften the EOS, reflected in an upward shift in $f_2$ and an earlier collapse of the merger \cite{arXiv:1809.01116}. This shift may be also caused by non-convex dynamics, involving expansive shocks and compressive rarefactions \cite{arXiv:2401.06849}.
The EOS model also dictates the maximum mass a cold, spherical, non-rotating neutron star can support before collapsing to a black hole, known as the  Tolman–Oppenheimer–Volkoff (TOV) limit. Observational constraints, combined with QUR, set a lower limit for this mass to $M_{TOV}\approx 2.2 M_{\odot}$, function of the EOS model \cite{arXiv:1711.00314, arXiv:1908.01012}. 
However, the BNS merger remnant is hot, highly-deformed, and spinning very fast, thus it can temporarily resist the collapse up to a threshold mass $M_{th} = k_{th} M_{TOV}$, where $k_{th} \approx 1.4$ \cite{arXiv:1901.09977}. When the mass $M$ of a BNS system is above this total mass, the merger remnant undergoes a prompt collapse into a black hole, while if it is below, it forms a massive neutron star remnant.
If the remnant does not collapse promptly, it can form a HyperMassive Neutron Star (HMNS) supported by differential rotation and thermal pressure -- this phase is dominated by GW emission and lasts only for tens of \texttt{ms} before the collapse to a black hole. 
If the remnant does not collapse during this early time, it transitions into the viscous phase, that can last up to several seconds, after which the star either collapses to a black hole if is supermassive, or it cools down and forms a stable neutron star when its mass is lower than $M_{TOV}$ \cite{arXiv:1703.10303}.    

It is also expected for the correlation described by eq.(\ref{eq:QUR}) to weaken with mass ratio, because for binaries with masses differing for more than $20\%$, the merger dynamics change qualitatively, the lower mass star being tidally disrupted and sheared apart before the merger. This causes significant mass to be ejected, even if the remnant undergoes prompt collapse \cite{arXiv:2003.06015, arXiv:2112.05864}. Moreover, unequal-mass BNS systems will merge faster, will have lower merger amplitude, and the modulations in the post-merger transient frequency will be suppressed, due to tidal disruption in the inspiral.
For mass ratios higher than $0.80$, the merger dynamics are similar to the equal-mass case, less material being ejected, primarily during the post-merger bounce of the cores \cite{arXiv:1908.02350}.
It was shown though that the mass loss during post-merger is $\le 0.2 M_{\odot}$ and although this decreases the tidal deformability, it has a small impact on QUR, creating a slight shift in $f_2$ \cite{arXiv:2106.08023}. 

\section{\label{sec:numerics} Numerical Techniques} 

We stated that NR codes simulating BNS mergers contain complex multiphysics and sophisticated numerical methods. 
Let us look at this problem in more detail.
The first step in the simulation of a BNS system is to prescribe the initial data for the binary.  
This is generated by solving the Tolman–Oppenheimer–Volkoff (TOV) equations, which describe the hydrostatic equilibrium inside the neutron stars. Those equations provide a solution to Einstein's equations of GR, $G_{\mu \nu} = 8 \pi T_{\mu \nu}$, for the energy-momentum tensor of a perfect fluid model: 
$T_{\mu \nu} = (\epsilon + P)u_{\mu} u_{\nu} + P g_{\mu \nu},$ 
where $\epsilon$ denotes the energy density, $P$ the pressure, $u_{\mu}$ the fluid 4-velocity, and $g_{\mu \nu}$ the spacetime metric.
To close the TOV equations, we must choose an EOS, that informs how a local metric suffers quadrupole deformations due to the companion.
This EOS is a thermodynamic equation describing the matter inside the neutron star by the relation between pressure, density and temperature: $P = P(\rho, T)$.
However, this EOS is still largely unknown, and it is usually phenomenologically modeled by parameters tuned to specific high-density configurations, informed by nuclear physics \cite{arXiv:1110.4442, arXiv:1207.2184, arXiv:1412.3240}. 
One common technique is to parametrize consecutive interior layers of a cold neutron star with a series of polytropes:
\begin{equation}
\label{eq:cold_eos}
    P_{cold} = K_i\rho^{\Gamma_i},~~
    \epsilon_{cold} = \epsilon_i + \frac{K_i}{\Gamma_i -1}\rho^{\Gamma_i-1}, ~~
    0<c_s^2 = \frac{d P_{cold}}{d \epsilon _{cold}} = \Gamma_i \rho <1,
\end{equation}
for a succession of density intervals $[\rho_i, \rho_{i+1}]$, where $K_i$ are polytropic constants that change with the stiffness of consecutive layers, $\Gamma_i$ are adiabatic indices, and $c_s$ is adiabatic speed of sound. 
These parameters must be chosen to ensure the pressure is monotonically increasing across the jumps in the energy density, corresponding to the transition between layers \cite{arXiv:0812.2163}. Additionally, since the density profile decreases outward, $c_s^2$ is also monotonically decreasing with radius.
To accurately model the neutron star interior, several polytropes can be used, that can describe phase transitions as well.
This formalism is implemented in initial data numerical codes that solve the TOV equations to compute the equilibrium configuration for a given binary neutron star system orbiting at a given separation.
Some of the most well known and used codes for BNS initial data are \texttt{LORENE} \cite{arXiv:gr-qc/0007028, arXiv:gr-qc/0103041, arXiv:gr-qc/0207098, arXiv:gr-qc/0309045}, \texttt{KADATH} \cite{arXiv:0909.1228}, \texttt{SGRID} \cite{arXiv:1209.5336}, \texttt{COCAL} \cite{arXiv:1502.05674}, and the more recently developed \texttt{FUKA} \cite{arXiv:2103.09911}. 
In the simulations we compare, thermal effects are incorporated using a hybrid approach that adds a thermal part to the piecewise polytropic EOS, described as an ideal gas EOS with an index $\Gamma_{th}$,
\begin{equation}
P = P_{cold} + (\Gamma_{th} -1) \rho (\epsilon -\epsilon_{cold}). 
\end{equation}
Another way to incorporate thermal effects into the EOS is to use Tabulated finite-Temperature EOS (TTEOS) as opposed to the Piecewise Polytropic EOS (PPEOS) model described above, with the added thermal part. TTEOS are indeed more accurate in modeling thermal effects and microphysics during the evolution, however they require interpolation of large data tables during simulations, which can introduce numerical errors and are computationally intensive \cite{arXiv:1006.3315}.

Once the initial data is provided, the numerical algorithms for evolving the BNS merger begin by prescribing suitable boundary conditions to the computational domain. Then the discretized equations describing the dynamics of matter, gravitational and electromagnetic fields, are solved in time using an appropriate numerical integration scheme such as $4^{th}$-order Runge-Kutta (RK4) or predictor-corrector methods, to ensure stability and accuracy during the evolution \cite{arXiv:1203.6443}.
  
The discretization of Einstein's field equations follows either the Baumgarte-Shapiro-Shibata-Nakamura (BSSN) prescription \cite{Shibata:1995, arXiv:gr-qc/9810065}, or the Generalized Harmonic (GH) formulation \cite{arXiv:gr-qc/0407110, arXiv:gr-qc/0512093}.
The relativistic MHD equations are represented by conservation laws for the relevant fluid and electromagnetic field variables, including the conservation of stress-energy tensor $\nabla_{\mu}T^{\mu \nu} = 0$ and the continuity equation for mass conservation $\nabla_{\mu}(\rho_b u^{\mu}) = 0$, where $\rho_b$ is the rest mass density.
Their discretization follows an involved path, because they must be rewritten in the flux-conservative form:
\begin{equation}
    \partial_{t} \mathbf C+ {\nabla}_{i}\mathbf F^{i} =\mathbf S,
    \label{eq:flux_evol}
\end{equation}
where $\mathbf C$ is the vector of conservative variables, $\mathbf F^{i}$ is the flux vector along direction $i$, and $\mathbf S$ is the vector of source terms. These vectors are functions of the primitive variables $\mathbf P$, which include the physical quantities describing the matter and electromagnetic field variables,  such as density, velocity, pressure, electron fraction, and magnetic field.
The spacetime metric is coupled to the (M)HD component of the code through the stress-energy tensor.
Other relevant physics, such as Maxwell’s equations, lepton number conservation and neutrino effects, are also included in the numerical algorithm.

Accurately evolving these complex equations during BNS merger simulations requires sophisticated numerical techniques capable of handling relativistic shocks and metric singularities.
Below, we present the widely used schematic numerical algorithm applied to the MHD variables at the start of a BNS simulation.
\begin{enumerate}
\item
First, initial data is provided at all the grid points for the primitive variables $\mathbf P$.
\item
These variables are converted into conservative variables $\mathbf C$ using algebraic formulas derived from the MHD equations. 
\item
An interpolator fills the boundary regions, ghost zones, and different levels of grid refinement necessary for the finite difference algorithm.
\item The flux terms $\mathbf F^{i}$ are evaluated as follows:
\begin{itemize}
\item
Calculate the primitive variables at the interfaces between grid cells using high-resolution methods, such as the Piecewise Parabolic Method (PPM) \cite{Colella:1982}.
\item
Convert the primitive variables at cell interfaces into conservative variables.
\item
Apply a High-Resolution Shock-Capturing (HRSC) scheme \cite{Kurganov:2000} to calculate the fluxes of the conservative variables across these interfaces.
\item
Compute the net fluxes across the grid interfaces from the fluxes on cell interfaces with an approximate Riemann solver, such as the Harten-Lax-van Leer (HLL) approach \cite{Harten:2006}.
\end{itemize}
\item
Calculate the sum of the spatial derivatives of the computed fluxes ${\nabla}_{i}\mathbf F^{i}$.
\item
Compute the source terms $\mathbf S$ necessary for the evolution of conservative variables, to account for neutrino radiation transport and microphysics \cite{arXiv:1703.10191, arXiv:2209.02538}.
\end{enumerate}

Now, the GRMHD equations (\ref{eq:flux_evol}) are ready to be integrated over time to evolve the conservative variables to the next timestep, based on the calculated net fluxes and source terms. 
After the time integration, the code must convert the updated conservative variables back into primitive variables. 
This process involves solving a non-linear system, and it is not always straightforward. Indeed, recovering the primitive variables from the conservatives is one of the most computationally challenging aspects of BNS simulations \cite{arXiv:1712.07538}. 
To solve the non-linear equations that define the primitive variables in terms of the conservative ones, a root-finding algorithm is employed, such as the Newton-Raphson method or the Noble \cite{arXiv:astro-ph/0512420} solver. 
Newton-Raphson is the most widely used method and requires an initial guess, which is iteratively refined by evaluating the Jacobian matrix and using its inverse for the function whose root is wanted.
The Noble method is specifically designed to handle highly relativistic flows, and implements a robust algorithm to ensure the accurate recovery of primitive variables, even under extreme conditions. This is important for simulations involving strong gravitational fields, such as those near black holes. We provide below a short overview of the approach:
\begin{enumerate}
\item
The solver starts with an initial guess for the primitive variables, based on the values from the previous timestep. 
\item
Next, non-linear equations derived from the conservation laws and the specific EOS used, relate the computed conservative variables to the guessed primitive variables.
\item
Then, the Jacobian matrix of these equations is calculated, and, if necessary, numerical approximations of its inverse are used, to increases robustness and to effectively handle highly non-linear terms.
\item
Lastly, the system of non-linear equations is solved with Newton-Raphson-like iteration, which involves updating the guess of the primitive variables by solving a linear system.
\item
After each iteration, the algorithm checks whether the changes in the primitive variables are below a threshold, the residuals of the equations meet a specified tolerance, and the solution converges.
\item
The step size may be adjusted to ensure stability and prevent non-physical values, such as negative pressures or densities, and superluminal speeds, especially near shocks or discontinuities.
\item
If the solver fails to converge, fallback strategies are triggered, such as using different initial guesses, switching to a more robust but computationally expensive method, or simplifying the equations temporarily to find a stable solution.
\end{enumerate}

During each timestep, this process is repeated to update the primitive variables based on the newly evolved conservative variables.
After the primitive variables are recovered and the physical constraints are reinforced, the conservative variables are updated at all grid points, in preparation for the next timestep.

\section{\label{sec:codes} Codes Specification}

Although numerous BNS simulations have been reported in the literature, our focus is on the small selection of codes that make their simulations publicly available or accessible upon request. Below, we provide a brief overview of the codes used in our analysis.

\begin{enumerate}
\item
{\bf SimulAtor for Compact objects in Relativistic Astrophysics (\texttt{SACRA})}

The \texttt{SACRA}
code \cite{arXiv:0806.4007, arXiv:1708.08926, arXiv:1711.02093, arXiv:1907.03790} implements Einstein's equations in the BSSN formalism \cite{Shibata:1995, arXiv:gr-qc/9810065} with moving punctures \cite{arXiv:gr-qc/0605030}, and enhanced with a Z4c constraint-propagation prescription \cite{arXiv:1212.2901}. 
It uses a $4^{th}$-order spatial finite differencing scheme on adaptive mesh refinement (AMR) \cite{arXiv:gr-qc/0610128}, with six buffer zones filled with a $5^{th}$-order Lagrangian interpolation and a $4^{th}$-order RK scheme for time integration. 
The hydrodynamic equations are solved in the flux-conservative form using the HRSC scheme and a PPM cell reconstruction, and includes also neutrino radiation transport \cite{arXiv:1703.10191}.
The waveform data, publicly available online \href{https://www2.yukawa.kyoto-u.ac.jp/~nr_kyoto/SACRA_PUB/catalog.html}{\cite{SACRA}}, includes 46 irrotational binary systems with six grid resolutions for each model, six different EOS, 
six mass ratios, binary total mass within $[2.5,2.74] M_{\odot}$, and inspiraling about $15$ orbits before the merger. 
The initial data are constructed with \texttt{LORENE} \cite{LORENE}, with eccentricity reduced to $10^{-3}$ \cite{arXiv:1405.6207}. 
A thermal part is added to the cold EOS during simulation to capture the shock heating effect around merger.
The finest reported grid refinement is between $64 \texttt{m}$ and $85 \texttt{m}$, depending on the system and the EOS considered.

\item
{\bf Bi-functional Adaptive Mesh (\texttt{BAM})}

The \texttt{BAM} code \cite{arXiv:gr-qc/0610128, arXiv:0706.0740, arXiv:1104.4751, arXiv:1504.01266}  evolves the Einstein's equations with a Z4c scheme \cite{arXiv:0912.2920, arXiv:1212.2901, arXiv:2109.04063} along with the moving puncture formalism \cite{arXiv:gr-qc/0605030} and a $4^{th}$ RK time integrator.
The spatial discretization is $4^{th}$-order accurate on a conservative AMR grid with seven refinement levels, three out of which are dynamically moving to follow the motion of each neutron star, and a multi-patch “cubed-sphere” for the wave zone. 
The GRMHD equations are solved in conservative form, using an HRSC method \cite{arXiv:1604.07999}, with primitive reconstruction performed using a fifth order WENOZ scheme \cite{Borges:2008}.
This code is part of the Computational Relativity collaboration \cite{arXiv:1806.01625, arXiv:2210.16366}, that maintains the CoRe database \cite{CoRe}, the largest release of public waveforms from BNS systems simulations, including a total of $254$ configurations. 
Out of these, $147$ simulations are obtained with the \texttt{BAM} code, for $11$ different EOS configurations and spanning a large range of masses, initial frequencies, spins, and eccentricities. 
BNS initial data are constructed with \texttt{SGRID} \cite{arXiv:1209.5336} and \texttt{LORENE} \cite{LORENE}, and a thermal part is added to the cold EOS during evolution. 
Most waveforms come from quasi-circular mergers with eccentricities of $e\approx 10^{-2}$.
A subset of $13$ configurations have eccentricity reduced to $10^{-3}$. 
The maximum reported resolution is approximately $88 \texttt{m}$, with overall second-order convergence during the inspiral and first-order convergence around the merger, continuing into the post-merger phase.

\item 
{\bf Templated-Hydrodynamics Code (\texttt{THC})}

The \texttt{THC}, or \texttt{WhiskyTHC} code \cite{arXiv:1306.6052, arXiv:1312.5004} represents the extension of the Newtonian \texttt{THC} \cite{arXiv:1206.6502} to GR, combined with a finite-volume MHD code \cite{arXiv:gr-qc/0403029, arXiv:1004.3849}.
This code evolves the spacetime with the BSSN formulation of the Einstein's equations \cite{Shibata:1995, arXiv:gr-qc/9810065}, using the Carpet AMR driver \cite{arXiv:gr-qc/0310042} and the \texttt{McLachlan} solver \cite{arXiv:1305.5299} within the \texttt{EinsteinToolkit} \cite{ETK} software for relativistic astrophysics.  
It implements a $4^{th}$-order finite-difference scheme for spatial discretization and $4^{th}$-order RK for time evolution.
The GRMHD equations are solved in conservative form by high-order finite-differencing central schemes \cite{Kurganov:2000, arXiv:1502.00551} that allow the inclusion of microphysics in the simulation, such as neutrinos transport. 
Initial data are calculated with either LORENE \cite{LORENE} or SGRID \cite{arXiv:1209.5336}.
This code is also part of the CoRe collaboration, contributing with $107$ simulations to the database, across nine different EOS configurations.
\texttt{THC} employs bitant symmetry, reports a maximum resolution around $118 \texttt{m}$, and self-convergence up to to $3^{rd}$-order \cite{arXiv:1306.6052}.

\item {\bf  \texttt{Whisky}}

The \texttt{Whisky} \cite{Whisky} code \cite{arXiv:1412.3240, arXiv:1604.00246} is built within the framework of the \texttt{EinsteinToolkit} \cite{ETK} as well, using the Carpet AMR driver and the \texttt{McLachlan} $4^{th}$-order finite-differencing code to solve Einstein's equations. The code evolves the GRMHD equations by HRSC schemes written in a flux-conservative form, with fluxes computed by a HLL approximate Riemann solver and primitive variables reconstructed at cell interfaces via PPM. 
Initial data is generated using the LORENE \cite{LORENE} code, and the cold piecewise polytrope EOS is combined with an ideal-fluid thermal part to form a hybrid EOS. 
The waveform data are available with permission at the Wiki link \cite{Wiki} and consists of 56 irrotational, mostly equal-mass neutron-star binary configurations, with six different EOS and ten distinct mass values between $2.4 M_{\odot}$ and $3.0 M_{\odot}$.
The reported convergence is approximately $2^{nd}$-order \cite{arXiv:1412.3240}, and resolution of about $220 \texttt{m}$ \cite{arXiv:1604.00246}.

\item {\bf Spectral Einstein Code Simulating eXtreme Spacetimes (\texttt{SpEC/SXS})} 

The \texttt{SpEC} \cite{SpEC} code solves Einstein's field equations in generalized harmonic coordinates \cite{arXiv:gr-qc/0407110} on a pseudo-spectral adaptive grid, using multi-domain spectral methods \cite{arXiv:0909.3557, arXiv:1405.3693}.
The GR hydrodynamics equations, written in conservative form, are evolved on a different, finite difference grid \cite{arXiv:0809.0002}. The time evolution uses a $3^{rd}$-order RK algorithm, and the two grids communicate at the end of each time step of the fluid evolution through polynomial interpolation.
It implements an original spectral prescription for the EOS \cite{arXiv:0804.3787}, shown to be smoother than tabulated or piecewise polytropic EOS \cite{arXiv:1908.05277}, constructed with the initial data \texttt{SPELLS} code \cite{arXiv:gr-qc/0202096} adapted for BNS systems. 
The GW signal is extracted at finite radii and extrapolated to null infinity using the method outlined in \cite{arXiv:0905.3177}.
The waveform data is limited to six configurations with two spectral EOS, chosen to match the GW170817 event \cite{arXiv:2307.03250}, and available upon request. 
The highest resolutions achieved are $174\texttt{m}-191\texttt{m}$, depending on the EOS. SpEC applies eccentricity reduction \cite{arXiv:gr-qc/0702106} and reports up to $3^{rd}$-order convergence in smooth regions, but only $1^{st}$-order convergence around shocks and discontinuities \cite{arXiv:1510.06398, arXiv:1604.00782}. 

\end{enumerate}

\section{\label{sec:methods}Methods}

Evaluating the performance and agreement between the codes summarized above is challenging due to the diverse methods employed. Different codes use a variety of formulations for the evolution of the Einstein equations, such as the Baumgarte-Shapiro-Shibata-Nakamura (BSSN), Z4C, or generalized harmonic formulations. They also adopt different prescriptions for the initial data, from \texttt{LORENE} and \texttt{SGRID} to \texttt{SPELLS}. 

As numerical codes rapidly evolve to incorporate increasingly complex multiphysics and continuously refine analytical models and numerical algorithms to improve accuracy and reliability, previously released simulation data quickly becomes outdated. This poses a challenge for building a large, long-term database of robust simulations, which waveform modelers must consider when relying on archival data.
Additionally, the specifics of the grid settings and the numerical techniques employed can differ—ranging from pseudo-spectral methods to finite difference or finite volume schemes, as can the methods for GW extrapolation.
Despite these challenges, we rely on known similarities between codes to distinguish between discrepancies due to physical modeling from those caused by numerical artifacts.
This approach also helps us evaluate how well the codes capture the underlying physics of BNS mergers. Gaining such insights is critical for enhancing the reliability of future simulations.

One commonality among all codes is that they compute the GW signal as the outgoing part of the Weyl curvature pseudo-scalar $\Psi_4$, related to the GW strain by $\ddot h = \Psi_4$. 
The documentation indicate that all codes compared here extract $\Psi_4$ data at a finite radius and compute the strain extrapolated to $\infty$ by integrating $\Psi_4$.
The GW signal is extracted in the wave zone on coordinate spheres by decomposing the metric perturbation about flat spacetime in complex spin-2 weighted spherical harmonics:
\begin{equation}
\ddot{h}(t) = \Psi_4(t) = \sum_{l=2}^{\infty} \sum_{m=-l}^l \Psi_4^{lm}(t) {}_{-2}{\cal Y}_{lm}(\theta, \psi)
\label{eq:waveform}
\end{equation}
Then, this equation is integrated to obtain the strain, following various prescriptions, one of the most used one being the fixed-frequency integration (FFI) method \cite{arXiv:1006.1632, arXiv:1012.0595}.

Achieving a clean, direct comparison between the waveforms obtained by different simulations is hindered by many factors, listed below. These factors can lead to measurable differences especially during the complex post-merger dynamics. 
\begin{itemize}
\item Differences in the physical assumptions, such as the influence of spins, magnetic fields, neutrino, ejected mass, or the choice of thermal adiabatic indices. 
\item Different techniques used to extrapolate the GW and the effect of the finite radius at which the signal is extracted on the accuracy of the waveform. 
\item
Variations in the initial separations between neutron stars and the number of orbits simulated before merger.
\item
Different choices in grid structure and refinement, boundary conditions, numerical dissipation and diffusion.
\item
Numerical artifacts in the initial data such as residual orbital eccentricity. 
\end{itemize}
While keeping in mind these differences between codes and simulations, we leave a detailed quantification of their individual contributions for future work. In this study, we focus on assessing the reliability of the simulated waveforms using two indicators: (1) the convergence of the simulated gravitational waveforms and (2) the reliability of the codes in reproducing quasi-universal relations. 
These two factors are relevant in providing insights on how to improve the accuracy of future simulations.

\subsection{\label{ssec:conv}New Approach to Code Convergence}

Convergence studies are essential for distinguishing numerical errors from physical effects in the waveform. By assessing convergence,  we can ensure that the numerical GW templates are of high-quality, a crucial factor for ensuring the accuracy of tidal deformability measurements from observed GWs.
The NR community reports robust and successful simulations performed with BNS codes, but the waveforms produced do not achieve the expected convergence at the nominal resolution. 
Codes do report $2^{nd}$-order convergence, independent of the particular details of the numerical simulation, almost up to the moment of the merger \cite{arXiv:2202.08839}.
However, achieving convergence across all stages of the evolution is challenging, especially during and after the merger, where it typically decreases to $1^{st}$-order convergence or becomes non-convergent, even when using higher-order hydrodynamic schemes.
Moreover, the convergence order shows irregular behavior, and the resolution generally proves insufficient to resolve sharp features, especially after the merger \cite{arXiv:1907.03790}.
For this reason, instead of relying on convergence measurements, numerical groups resort to comparing phase differences between resolutions, and to tracking the error, which for consistent results, should decrease with refinement. 
However, even the residual phase error, similar to the convergence order, shows irregular behavior, changing sign during the evolution, which increases the numerical noise \cite{arXiv:1604.00782}.
Additionally, simulations using different resolutions of the same initial data and evolution code tend to diverge over time as phase uncertainty increases during evolution, especially during the final cycles of the GWs, when the stars come into contact.
Furthermore, numerical dissipation due to insufficient grid resolutions can cause artificial losses of angular momentum and mass during evolution. These losses can mimic tidal effects and shift the merger time, making it hard to locate. 
To complicate the convergence analysis, more compact stars, associated with softer a EOS, require higher numerical resolution to achieve small phase errors. This makes it challenging to compare resolution effects for GWs across different EOS models, as each EOS may demand different resolutions for comparable accuracy. 
Convergence also proves more challenging for high spins and unequal-mass binaries, especially in the extreme mass ratio regime \cite{arXiv:2103.09911}.

Thus, achieving convergence in BNS simulations is a difficult task, especially in the highly non-linear region around merger, due to shock formation during collision and complex fluid dynamics in the remnant.  
Among the many factors that can introduce errors affecting the convergence, the most important are the truncation errors due to the choice of numerical approximation used in the hydrodynamic schemes, and the discretization errors due to the finite resolution of the computational domain \cite{arXiv:1806.01772}.

The commonly used method to calculate the convergence and to quantify the discretization errors in numerical simulations is to use the Richardson extrapolation.
This method is based on the assumption that the numerical solutions can be represented as Taylor series in terms of the grid size, with the first term of the expansion dominant:
\begin{equation}
F_e = F_r + \sum_{i=p}^{\infty} c_i h_r^i \approx c h_r^p,
\end{equation}
where $F_r$ is the numerical solution on the grid of size $h_r$, $F_e$ is the exact solution, $c_i$ are coefficients independent of the gridsize, and $p$ is the order of convergence.
Then, if the solution is known on three systematically refined grids of gridsizes: $h_c>h_m > h_f$ (coarse, medium and fine), it is possible to form a system of three equations, 
\begin{eqnarray}
F_e - F_c = c h_c^p, \\
F_e - F_m = c h_m^p,\\
F_e - F_f = c h_f^p,
\end{eqnarray}
and to solve for the three unknowns: $F_e$, the coefficient $c$ and the convergence order $p$.
Then, the convergence is estimated numerically solving the self-convergence formula given below, obtained from the above system of equations: 
\begin{equation}
\label{eq:method1}
 {\frac{F_m - F_f}{F_c - F_f}}  = \frac{h_f^p - h_m^p}{h_f^p - h_c^p} = \frac{1 - r_{mf}^p}{1- r_{cf}^p},
\end{equation}
where $r_{cf} = h_c/h_f$ and $r_{mf} = h_m/h_f$.
Note that if the grid ratio is kept constant $(r_{mf} = r_{cm} = r)$, then $r_{cf} = r^2$ and eq.(\ref{eq:method1}) simplifies to:
\begin{equation}
p = \frac{1}{\ln(r)} \ln \left ( {\frac{F_c - F_m}{F_m - F_f}} \right ) = \frac{1}{\ln(r)} \ln ( \cal R),
\end{equation}
where ${\cal R} = (F_c - F_m)/(F_m - F_f)$ is the convergence ratio.

However, this method is limited by its reliance on smooth, asymptotic numerical solutions and the assumption that the discretization error dominates all other errors, expecting a constant order of convergence that matches the accuracy of the discretization scheme.  In computational fluid dynamics problems, Richardson extrapolation will fail in the shock regions, where the error behavior is not monotone and the accuracy of the higher-order non-linear methods employed is lost \cite{errorWENO}. In this situation, smooth regions must be identified where convergence calculations could safely proceed \cite{richardson2}. 

The convergence type of a numerical solution can be classified based on the convergence ratio ${\cal R}$, as: monotonic convergence ($0<{\cal R} < 1$), monotonic divergence (${\cal R} > 1$), oscillatory convergence ($-1<{\cal R} < 0$), and oscillatory divergence (${\cal R} <-1$). 
Most studies consider only monotonic convergence and discard the other types of convergence, because they do not fit with the theory of Richardson extrapolation. 
Nevertheless, Richardson extrapolation can still be applied to problems with discontinuities, where the local convergence rate exhibits oscillatory or even divergent behavior, which reduces the order of accuracy from what would be expected from the discretization scheme \cite{richardson3}.

One such method, introduced in \cite{richardson1}, can be applied to solutions with oscillatory convergence and supports non-uniform grids. In this work, we apply this method for the first time to BNS simulations, enabling us to safely proceed with Richardson extrapolation and study oscillatory convergence. The main idea of the method is to take into account the sign of the convergence ratio:
\begin{equation}
s = \texttt{sign} (\cal R),
\label{eq:s}
\end{equation}
where negative values of $s$ indicate oscillatory convergence.
We include this sign in the Richardson extrapolation system of equations: 
\begin{eqnarray}
F_e - F_c = c h_c^p, \\
F_e - F_m = s c h_m^p,\\
F_e - F_f = c h_f^p.
\end{eqnarray}
and compute the convergence order $p$ by numerically solving the relation:
\begin{equation}
\label{eq:method2}
 {\frac{F_m - F_f}{F_c - F_f}}  = \frac{h_f^p - s h_m^p}{h_f^f - h_c^p} =\frac{1 - s r_{mf}^p}{1- r_{cf}^p}.
\end{equation}

We refer to the first approach, as expressed by eq.(\ref{eq:method1}), as the Monotonic Convergence (MC) and the second one, detailed in eq.(\ref{eq:method2}), as the Oscillatory Convergence (OC).
We apply both techniques to determine convergence.
It is important to note that the convergence factor calculation may fail when the differences $\epsilon_{cm} = F_c - F_m$, $\epsilon_{cf} = F_c - F_f$, or $\epsilon_{mf} = F_m - F_f$ approach zero. In cases where no solution is found, we substitute the local order of convergence with a corrective factor $p_{CF} = 0.5$. 
To account for the effects of oscillatory converging nodes, we then compute and report the average convergence using the absolute values of the local convergence order:
\begin{equation}
\label{eq:pave}
{\bar p} = \frac{1}{N}\sum_{i=1}^N |p_i|,
\end{equation}
where $N$ is the number of grid cells or nodes.
 We also estimate the deviation from the average convergence order over the entire domain with a formula used in computational fluid dynamics studies \cite{richardson3}:
\begin{equation}
\label{eq:delp}
 \Delta {\bar p} = \min \left ( \frac{1}{N}\sum_{i=1}^N \min(|{\bar p} - p_i|, 4{\bar p} ), 0.95 {\bar p}\right ).
\end{equation}

We calculate the average convergence and the deviation from the expected order of accuracy for the waveforms generated by the codes under study, provided that three solutions on  consecutive grids are available. 
Because it is not feasible to compute convergence for all data analyzed here, we focus on evaluating the convergence performance and accuracy for a set of representative EOS at three different mass ratios.

\subsection{Quasi-universal Relations Comparisons}

Establishing convergence is understandably a complex task, given the inherent challenges of maintaining code accuracy when dealing with shocks and discontinuities in BNS simulations. However, certain representative frequencies in the GW spectra have been shown to exhibit robust features, with a clear dependence on the $1/5$ power of the tidal deformability $\tilde \Lambda$ \cite{arXiv:1908.11418, arXiv:2205.09112, arXiv:2310.10728}. This motivates us to continue our analysis by assessing how well the simulated waveforms from the codes studied here predict quasi-universal relations (QURs) and to quantify any discrepancies.

We analyze the GW output from a large set of irrotational BNS systems, covering a wide range of effective tidal deformabilities, total mass, and mass ratios. From the GW data, we compute the three key frequencies $(f_{mrg}, f_{max}, f_2)$ for both the strain $h$ and the Weyl scalar $\Psi_4$.
Ideally, these frequencies should match between the strain and the Weyl scalar, however, in practice, they differ, due to numerical errors. For example, depending on the method used for the GW extraction, the strain can be affected by errors introduced through double integration and extrapolation.
We focus exclusively on the dominant $(l=2,m=2)$ mode of the waveform, as defined by eq.(\ref{eq:waveform}) and omit the indices for simplicity:
\begin{equation}
\Psi_{4,22}(t) \approx \Psi_{4}(t) = A_{\Psi_4}(t) \e^{-i  \phi_{\Psi_{4}}(t)}, 
\end{equation}
\begin{equation}
h_{22}(t) \approx h(t) = A_{h}(t) \e^{-i  \phi_{h}(t)},
\end{equation}
where $A_{\Psi_4}(t)$ is the amplitude of the Weyl scalar, $A_h(t)$ the amplitude of the strain, and $(\phi_{\Psi_{4}}(t), \phi_{h}(t))$ the phases of their respective waveforms.
These quantities are complex and are further decomposed in their real and imaginary parts, corresponding to the $+$ and $\times$ polarizations of the radiated signal: 
\begin{equation}
 \Psi_{4}(t) = \Re[\Psi_{4}(t)] - i \Im[\Psi_{4}(t)] = \Psi_{4,+}(t) - i \Psi_{4,\times}(t), 
 \label{eqLpsi4}
\end{equation}
\begin{equation}
h(t) = \Re[h(t)] - i \Im[h(t)] = h_{+}(t) - i h_{\times}(t).
\label{eq:h}
\end{equation}
For the reasons previously discussed, the phases of the strain and Weyl scalar differ and we extract them individually: 
\begin{equation}
\phi_{\Psi_4}(t) =  \arctan \frac{\Psi_{4,+}(t)}{\Psi_{4,\times}(t)},
\end{equation}
\begin{equation}
\phi_{h}(t) = \arctan \frac{h_{+}(t)}{h_{\times}(t)}.
\label{eq:phaseh}
\end{equation}
Before calculating the instantaneous frequency by taking the time derivative of the phase, we must first convert the time coordinate from code units to seconds by multiplying it by $u_t$. 
This conversion ensures the frequency, as calculated below, is in units of Hertz, enabling consistent comparison across different codes.
\begin{equation}
f_{\Psi_4}(t) = \frac{\omega_{\Psi_4}(t)}{2\pi} = \frac{1}{2\pi}\frac{d \phi_{\Psi_4}(t)}{dt},
\end{equation}
\begin{equation}
f_{h}(t) =\frac{\omega_h (t)}{2 \pi}= \frac{1}{2\pi}\frac{d \phi_h(t)}{dt}.
\end{equation}

The next step is to identify the key frequencies for comparison: the merger frequency $f_{mrg}$, the maximum frequency $f_{max}$, and the dominant $(l=2, m=2)$ mode frequency $f_2$. First, we define the merger as the point where the strain amplitude, $A_h(t) = \sqrt{h^2_{+}(t)+h^2_{\times}(t)}$, reaches its highest peak and the $\Psi_4$ amplitude, $A_{\Psi_4}(t) = \sqrt{\Psi^2_{4,+}(t)+\Psi^2_{4,\times}(t)}$ attains its first maximum.
After we identify the time of the merger, we extract the corresponding frequencies: $f_{h,mrg} =f_{h}(t_{h,mrg}) $ and $f_{\Psi_4,mrg} =f_{\Psi_4}(t_{\Psi_4,mrg})$ from the post-merger signal for both strain and $\Psi_4$.

Next, we track the post-merger evolution of the amplitude, identifying the time at which the strain amplitude reaches its lowest minimum and the Weyl scalar amplitude reaches its first minimum. Around this time, we collect the highest frequency, which corresponds to the moment when the stellar cores are at their closest approach, just before bouncing off. It is important to note that this is a transient phase, characterized by shocks and non-linear dynamics.
Due to limited resolution and the reported loss of simulations convergence in the post-merger, many codes struggle to accurately capture this frequency. 
This behavior may be caused also from the instantaneous frequency becoming ill-defined near the amplitude minimum.
As a result, it is not always identified correctly and can vary significantly, sometimes even having negative values.

Our final task is to calculate the dominant post-merger GW emission frequency, $f_{2}$. 
To this end, we construct the spectrogram of the signal to track the evolution of the effective amplitude, as defined by its Fourier transform. 
This represents the Power Spectral Density (PSD) that quantifies how the emitted GW energy is distributed across frequencies over time. This spectrogram simplifies the identification of the dominant monochromatic frequency at which the intensity of the signal is strongest for the longest duration.
Systems that collapse into a black hole can be identified by a sharp, asymptotic increase in this frequency followed, by a region without any dominant frequency.
 
We assess the performance of the codes by comparing their relative errors against empirical fits to the QURs and by examining the consistency of the fitting coefficients across different codes.
Systems with short post-merger data or those that collapse immediately after the merger are excluded from the analysis of the QUR fits for $f_2$, as this it is not distinctly observable.

For the first time, we examine the relationship between the transient time from merger to peak post-merger frequency $\Delta t_{f_{max},f_{mrg}} = t_{f_{max}} - t_{f_{mrg}}$ and the EOS model. Our analysis reveals that this interval also scales with tidal deformability, according to the following relationship:
\begin{equation}
\log(\Delta t_{f_{max},f_{mrg}}/M)=a+b\Lambda^{1/5}
\end{equation}
This time interval is particularly useful for capturing the transient phase immediately following the merger.
Accurately resolving this region is significantly challenging for numerical codes due to its dynamic nature and the presence of shocks. Therefore, monitoring this interval is important for assessing code performance.

\section{\label{sec:results}Results}

\subsection{Data Organization and Processing}

The first step in organizing the data is to establish a selection process to determine which GW signals to include in our comparison. We focus exclusively on irrotational binary systems with negligible eccentricity. For the \texttt{SACRA} code, we include all $46$ configurations at the three highest refinement levels. 
For the data in the CoRe database, which includes GW signals obtained with both \texttt{BAM} and \texttt{THC} codes, the selection process is more meticulous. We choose only irrotational, non-eccentric binary systems from this database, at a single extraction radius, selecting the last radius that offers the maximum available resolution runs. 
Then each simulation is carefully evaluated to identify inconsistencies and determine its inclusion based on the successful identification of key frequencies.
For the \texttt{Whisky} code, we use all available $59$ waveforms, which provide only the strain data at a single resolution. As a result, we cannot perform a convergence analysis on this data set.
Finally, the \texttt{SpEC} code includes only $2$ configurations at $3$ specific mass ratio, and we use all these $6$ datasets in our analysis.

We follow a naming convention to label the GW data: \texttt{Code\_EOS\_q\_M}, which provides key information about the code used to generate the data, the name of the EOS used, the mass ratio and the total mass of the BNS system. For example: $\texttt{BAM0126}\_\texttt{SLy}\_080\_275$ denotes the waveform obtained with the \texttt{BAM} code, using the $\texttt{SLy}$ EOS, for a BNS system with mass ratio $q = 0.80$ and total mass of $M = 2.75 M_{\odot}$. Note that for the codes in the \texttt{CoRe} database, the label also includes the run number for ease of identification.

\subsection{Code Comparison for Similar Tidal Deformabilities}

It is important to compare whether different codes yield the same key parameters, such as merger amplitude, frequency, and the maximum and dominant frequencies, for the same EOS, mass, and mass ratio. However, we do not have data that exactly matches these conditions across all five codes. As a workaround, we limit our selection to equal-mass binaries with the closest available tidal deformabilities and total mass.
We find that all codes contain data from simulations with a tidal deformability corresponding to the estimated value for the GW170817 event, with an average around $\overline{\tilde \Lambda} = 597.8$, a spread of only $0.72\%$ and a total mass $\overline{M} = 2.714 M_{\odot}$ with relative deviation of $0.87\%$.

When comparing simulations from different codes, the first challenge is the variability in the initial separation between neutron stars. This affects the number of orbits and influences the numerical error accumulated before merger, which introduces a dephasing between simulations due to changes in the accumulated GW phase.
Shorter simulations might miss early tidal effects in the inspiral, while longer ones risk a greater numerical error that can affect the phase evolution. Residual eccentricity from the initial data can also affect the simulation, complicating direct comparisons.

To enable a meaningful comparison and ensure that the time labels correspond to the same event, we identify the merger time as the reference event where all waveforms should align. This is defined as the point where the strain amplitude reaches its maximum. We then apply relative time and phase shifts to synchronize the waveforms at this event, by adjusting the waveforms such that the merger occurs at the same time and phase, $t_0 = 0$ and $\phi(t_0) = 0$ for all simulations.
This method allows for waveform comparison despite differences in the initial separations. 
Although these differences, along with residual eccentricity and numerical errors accumulated during the early inspiral, may introduce uncertainties, we assume that their impact on the calculated amplitude and frequency is relatively minor and unaffected by time shifts.

\begin{figure}[h]
    \centering
    \includegraphics[width=0.8\textwidth]{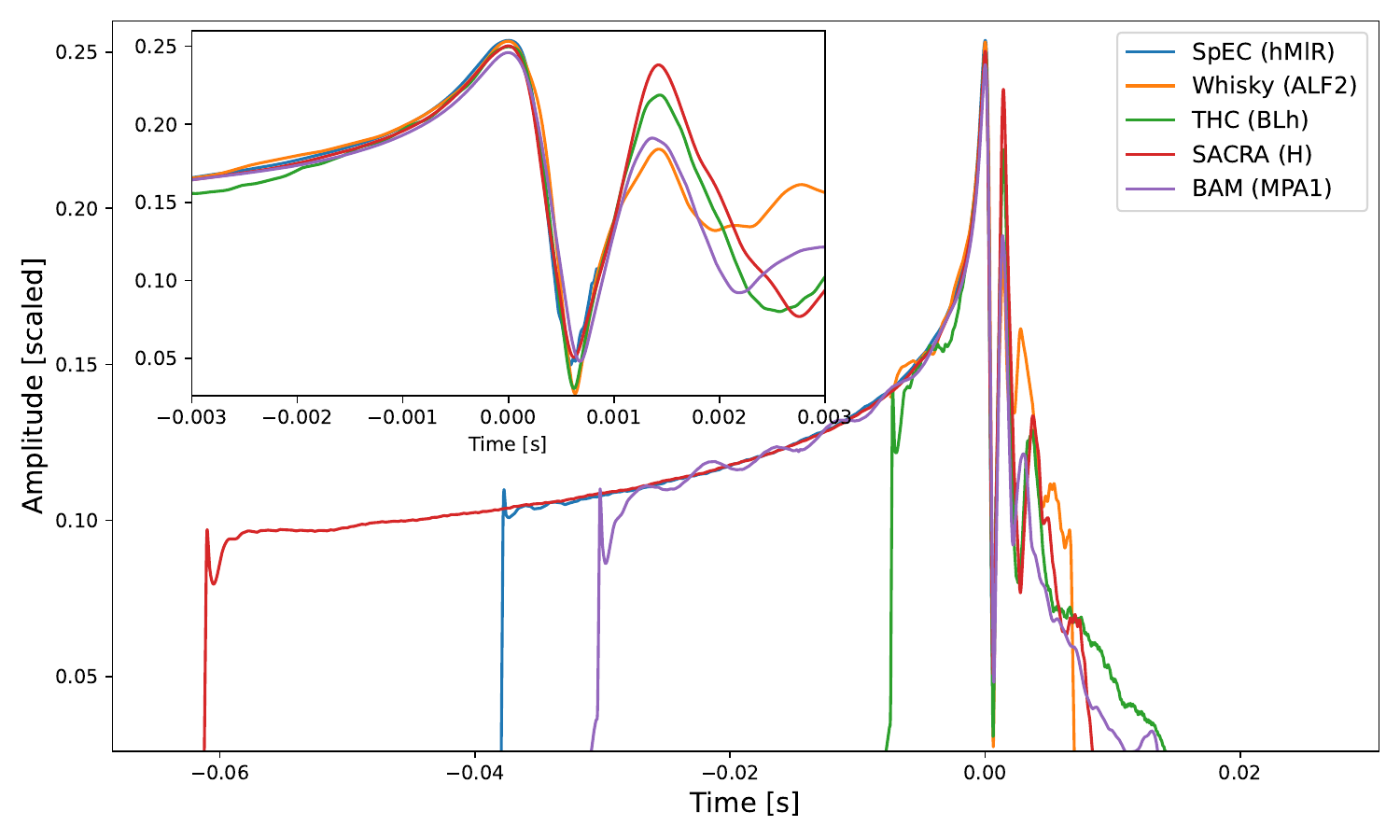}  
    \caption{Evolution of the GW rescaled strain amplitude $A_h/M$ for the five codes considered, with the inset zooming in on the merger region.}
    \label{fig:figure1}
\end{figure}

We illustrate the method for our comparison with the following steps. 
Initially, we convert the time coordinate from code units to seconds using the conversion factor $u_t$. 
For \texttt{SpEC}, the time coordinate requires additional multiplication by the total mass, while for \texttt{Whisky} and \texttt{SACRA}, the strain needs to be rescaled by dividing with the total mass. We then calculate the amplitude, and adjust the time coordinate so that $t_0 =0$ corresponds to the peak amplitude. We present these processed waveforms in Figure \ref{fig:figure1}, with code names and EOS  listed in the legend in increasing order of tidal deformability.
 
All codes successfully capture the variation in amplitude within the transient post-merger region, characterized by a sharp minimum immediately following the merger and then a second, less pronounced maximum. However, the concordance in amplitude across codes diminishes beyond the first minimum.
\texttt{BAM}, and to a lesser extent \texttt{SpEC}, amplitudes show oscillations in the early inspiral, likely due to residual eccentricity in the initial data. \texttt{Whisky} and \texttt{THC} exhibit early noise, taking longer to dissipate initial junk radiation.
Moreover, initial data errors also introduce fluid dynamics effects that excite fluid motion, which does not radiate away on a fast timescale. \texttt{SACRA} achieves the smoothest amplitude, and has the longest evolution.

\begin{figure}[h]
    \centering
    \includegraphics[width=0.8\textwidth]{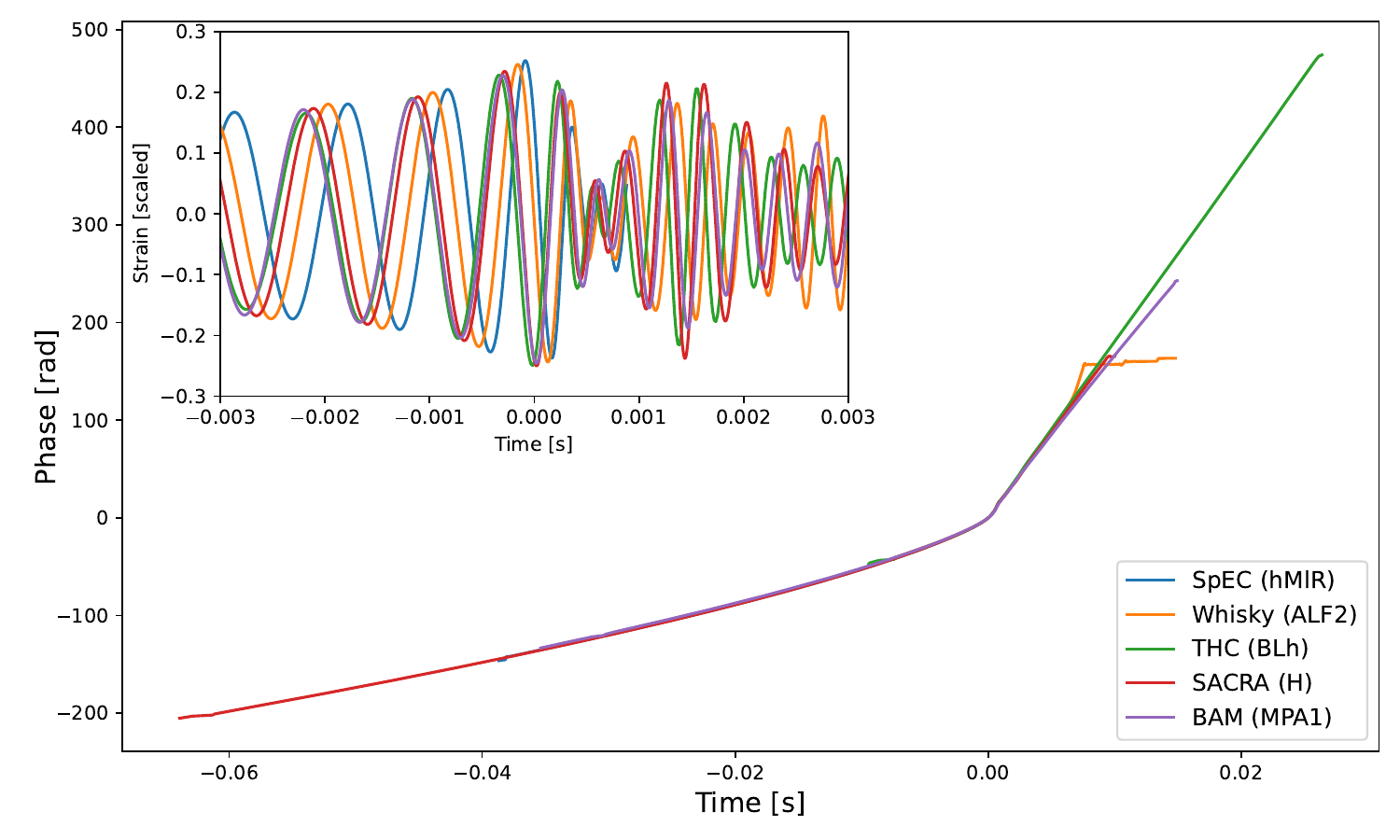}  
    \caption{Evolution of the GW phase for the five codes considered, with the inset containing the real part of the strain, zoomed in around the merger region.}
    \label{fig:figure2}
\end{figure}

Next, we calculate the phase for the strain, using eq.(\ref{eq:phaseh}). 
Figure \ref{fig:figure2} shows the phase comparison over the entire duration available for the data compared, with an inset focusing on the strain around merger. 
A good concordance in phase is observed between \texttt{THC}, \texttt{SACRA} and \texttt{BAM} data, while \texttt{SpEC} and \texttt{Whisky} strains seem to fall slightly out of sync.
We also examine the phase at the earliest and latest common times to evaluate inter-code differences.
The latest start time is from \texttt{Whisky}, which begins at approximately $-7 \texttt{ms}$. 
At this point, the phase values range between $\theta\in(-40.1, -39.05)\texttt{rad}$, with the largest phase difference less than $1.0\texttt{rad}$ observed between \texttt{SpEC} and \texttt{BAM}, consistent with the corresponding tidal deformabilities.
The earliest common end time is from \texttt{SpEC}, which ends shortly after merger, around $ 0.8 \texttt{ms}$, where $\theta\in(14.15, 15.6)\texttt{rad}$. 
Here, the largest phase difference, $\Delta \theta\approx 1.45\texttt{rad}$, occurs between \texttt{BAM} and \texttt{THC}. This phase divergence is also visible in the inset, where the strains fall out of sync in the post-merger.
After the merger, no evident dependence on tidal deformability is observed, as phase differences between codes are large enough to obscure the subtle variations in the chosen EOS.

\begin{figure}[h]
    \centering
    \includegraphics[width=0.9\textwidth]{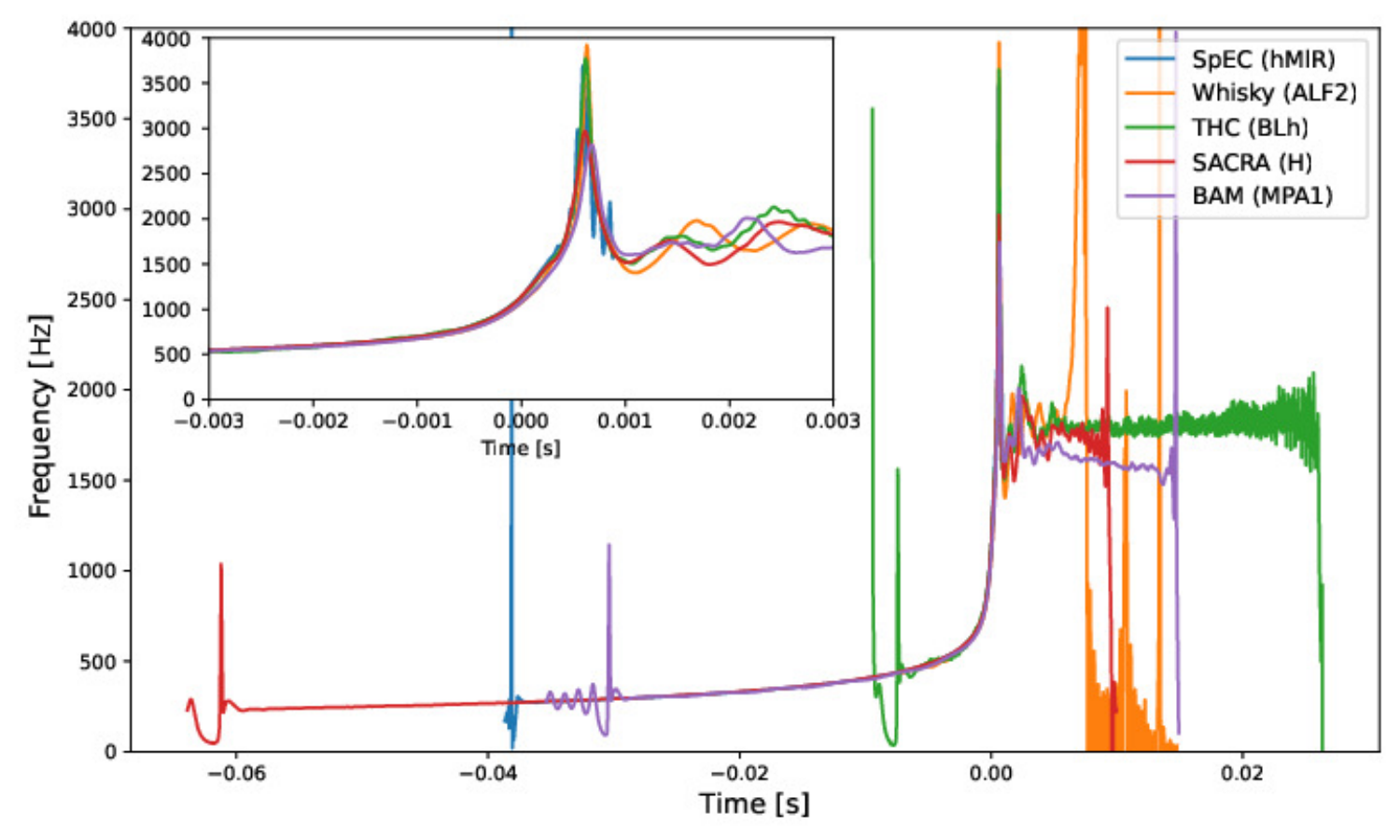}  
    \caption{Evolution of the GW frequency for the five codes considered, with the inset zooming in on the merger region.}
    \label{fig:figure3}
\end{figure}

With the phase determined, we calculate and extract representative values for the merger and maximum frequencies. Figure~\ref{fig:figure3} shows a comparison of the GW frequency evolution, with an inset focusing on the merger region. We observe that, after an initial spike caused by junk radiation, the frequencies of the five codes align well during the inspiral and up to the frequency peak in the early post-merger phase. Beyond this point, data for \texttt{SpEC} is unavailable, and \texttt{Whisky} shows noticeably unphysical spikes.

\begin{figure}[h]
    \centering
    \includegraphics[width=0.9\textwidth]{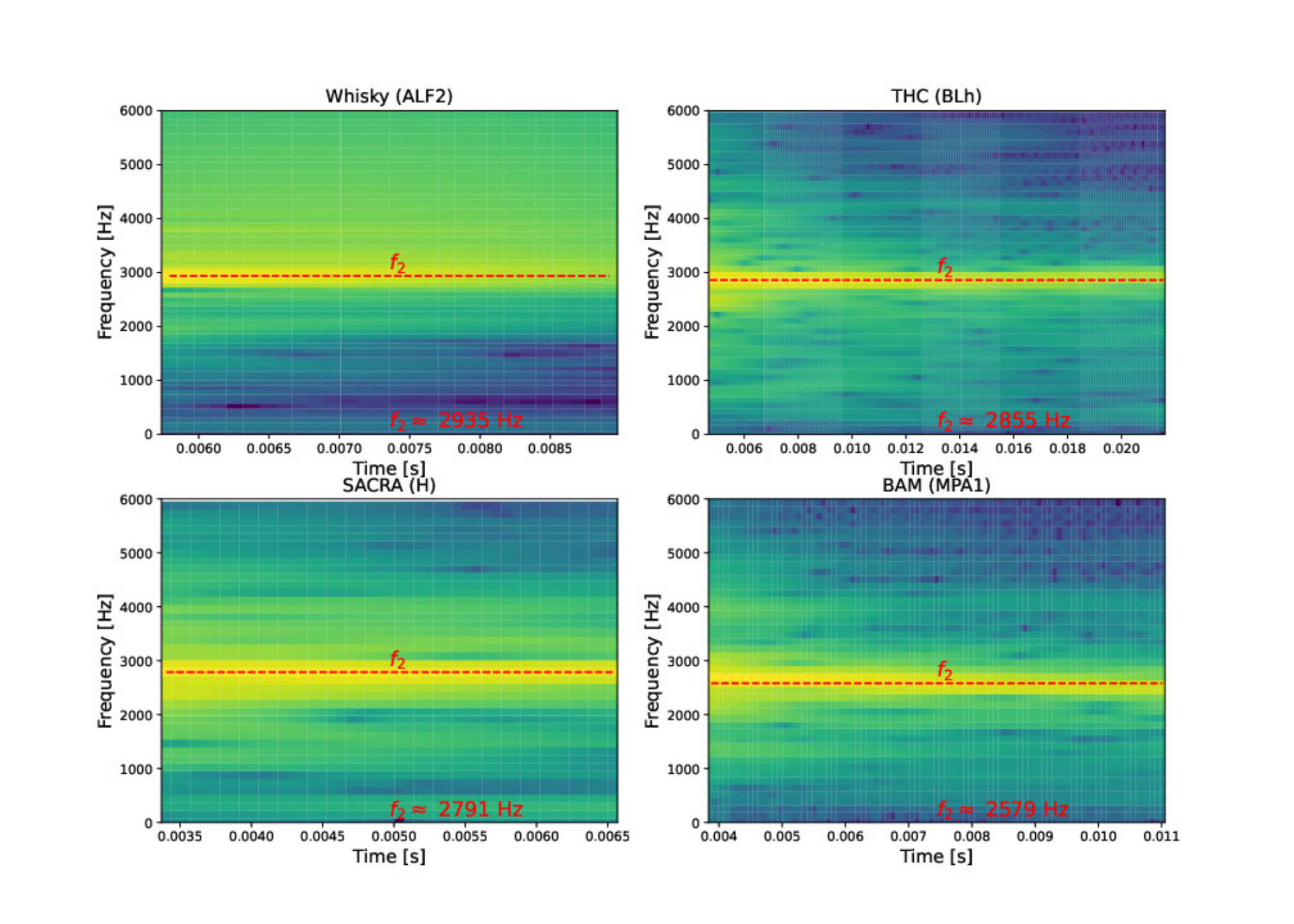}  
    \caption{Spectrograms of the post-merger strain for four of the five codes considered, with a horizontal dashed line marking the dominant frequency.}
    \label{fig:figure4}
\end{figure}

Finally, we construct the spectrogram for the strain and extract the dominant frequency $f_2$, which corresponds to the $(l = m =2)$ fundamental mode of the HMNS formed after merger.
This is done by identifying the frequency of the maximum total power output.
The post-merger spectrograms are shown in Figure \ref{fig:figure4}, with $f_2$ indicated by a horizontal line.
Data for \texttt{SpEC} is unavailable, as it ends shortly after the merger.

Table \ref{tab:tab1} presents the mean and standard deviations for the merger, maximum, and dominant frequencies. The relative deviation does not directly measure code performance, but it highlights the ability to resolve minor differences in tidal deformability during and after the merger. As expected, the error is largest in the maximum frequency.
\begin{table}[h]
\centering
\caption{\label{tab:tab1}Mean and values percentage errors for the compared parameters.}
\begin{tabular}{@{}llll}
\br
$\overline{f_{mrg}}[\texttt{kHz}]$&$\overline{f_{max}}[\texttt{kHz}]$&$\overline{f_{2}}[\texttt{kHz}]$\\
\mr
$1.112 \pm 0.064~(2.11\%)$ & $3.438 \pm 1.29 ~(13.91\%)$ & $2.799 \pm 0.44 ~(5.87\%)$\\
\br
\end{tabular}
\end{table}

In Table \ref{tab:tab2} we present the selected key parameters, along with the corresponding tidal deformabilities for each code.
We observe for example that the decrease in merger amplitude appears to correlate with an increase in tidal deformability, reflecting the known effect that a softer EOS leads to a more energetic merger. This trend is evident even within the very narrow spread of  $0.56\%$ in merger amplitudes, centered around a mean value of $\overline{A_{mrg}/M} = 0.251$, with the exception of \texttt{SACRA} data.
We observe the strongest correlation in the dominant postmerger frequencies, which consistently show an inverse relationship with tidal deformability. However, this correlation is less consistent for the merger and maximum frequency, and for the time intervals between the merger and maximum frequencies. This finding will be validated by our QUR analysis.
\begin{table}[h]
\centering
\caption{\label{tab:tab2}Selected data and key parameters for equal-mass binaries comparison.}
\begin{tabular}{@{}lllllll}
\br
Data&$\tilde \Lambda$&$A_{mrg}$&$f_{mrg}[\texttt{kHz}]$&$f_{max}[\texttt{kHz}]$&$f_{2} [\texttt{kHz}]$&$\Delta t_{f_{max}}[\texttt{ms}]$\\
\mr
$\texttt{SpEC}\_\texttt{hMlR}\_100\_272$        &$588$&$0.2538$&$1.137$&$3.696$&$N/A$&$0.2172$\\
$\texttt{Whisky}\_\texttt{ALF2}\_100\_280$      &$591$&$0.2533$&$1.107$&$4.045$&$3.028$&$0.2246$\\
$\texttt{THC0100}\_\texttt{BLh}\_100\_266$    &$594$&$0.2498$&$1.110$&$3.696$&$2.827$&$0.2311$\\
$\texttt{SACRA}\_\texttt{H}\_100\_270$           &$607$&$0.2503$&$1.134$&$2.951$&$2.777$&$0.2269$\\
$\texttt{BAM0058}\_\texttt{MPA1}\_100\_270$ &$609$&$0.2460$&$1.071$&$2.801$&$2.566$&$0.2513$\\
\br
\end{tabular}
\end{table}

\subsection{Convergence for Selected Waveforms}
We now turn to the challenging task of analyzing code convergence, a well-known issue in BNS simulations.  Ideally, we would use the previously selected data with similar tidal deformability values to assess convergence; however, not all codes provide three refinement levels required for this calculation. 
Convergence studies cannot be performed on data from \texttt{Whisky} and \texttt{SpEC}, due to insufficient resolutions availability.

For these reasons, we have expanded our data selection to include additional tidal deformabilities and mass ratios, for a total of 15 configurations. This selection provides a representative dataset for assessing convergence across a variety of EOS models and mass ratios, for configurations with total mass between $2.7 M_{\odot}$ and $2.8 M_{\odot}$. 
To synchronize the time steps and ensure they are coincident across different levels of refinement, all datasets require resampling. We achieve this through $4^{th}$ order univariate spline interpolation, using a timestep equal to the largest one found in any dataset at all three resolutions.

We iteratively solve eqs.(\ref{eq:method1}) and (\ref{eq:method2}) with an initial guess for $p$, specifically the estimated order of $p=2$ commonly reported in BNS code convergence.
We calculate the average convergence order and deviation for the entire time domain, including the post-merger phase, for the real part of the evolved variable $\Psi_4$, and present our results in Table \ref{tab:tab3}. The subscripts  $MC$ and $OC$ represent the convergence order and deviation calculated with the two methods presented in Section \ref{ssec:conv}.
We find that the convergence orders obtained with the two methods are similar, though \texttt{SACRA} favors the OC method.
This code shows a distinctly higher convergence for binaries with equal-mass and with mass ratio close to $1$. However, the convergence drops sharply for mass ratio $q \le 0.8$. On the other hand, \texttt{BAM}, while displaying a lower overall convergence rate, has minimal variation in convergence order across different mass ratios. \texttt{THC} demonstrates the lowest convergence order among the three, but shows consistent behavior across mass-ratios as well. For \texttt{SACRA} and \texttt{BAM}, where two different EOS were tested, the results 
do not show a significant influence of tidal deformability on the convergence order.

\begin{table}[h]
\centering
\caption{\label{tab:tab3}Monotonic (MC) and oscillatory (OC) code convergence calculation.}
\begin{tabular}{@{}llllllll}
\br
Data&$\tilde \Lambda$&Mass& $q$ & $\bar p_{MC} \pm \Delta \bar p_{MC}$ & $\bar p_{OC} \pm \Delta \bar p_{OC}$\\
\mr
$\texttt{SACRA}\_\texttt{H}\_100\_270$&$607$&$2.70$&$1.00$&$2.02 \pm 1.52$&$3.42 \pm 2.08$\\
$\texttt{SACRA}\_\texttt{H}\_086\_271$&$605$&$2.71$&$0.86$&$1.83 \pm 1.69$&$2.05 \pm 1.75$\\
$\texttt{SACRA}\_\texttt{H}\_080\_272$&$604$&$2.72$&$0.80$&$0.85 \pm 0.53$&$0.89 \pm 0.59$\\
$\texttt{SACRA}\_\texttt{HB}\_135\_270$&$422$&$2.70$&$1.00$&$2.09 \pm 1.62$&$3.38 \pm 1.95$\\
$\texttt{SACRA}\_\texttt{HB}\_086\_271$&$423$&$2.71$&$0.86$&$2.14 \pm 1.90$&$2.78 \pm 2.32$\\
$\texttt{SACRA}\_\texttt{HB}\_080\_272$&$422$&$2.72$&$0.80$&$0.74 \pm 0.36$&$0.75 \pm 0.39$\\
$\texttt{BAM0070}\_\texttt{MS1b}\_100\_275$&$1389$&$2.75$&$1.00$&$1.07 \pm 0.84$&$0.97 \pm 0.68$\\
$\texttt{BAM0089}\_\texttt{MS1b}\_080\_275$&$1420$&$2.75$&$0.80$&$0.97 \pm 0.73$&$0.92 \pm 0.64$\\
$\texttt{BAM0091}\_\texttt{MS1b}\_066\_275$&$1484$&$2.75$&$0.66$&$0.90 \pm 0.64$&$0.93 \pm 0.68$\\
$\texttt{BAM0120}\_\texttt{SLy}\_100\_275$&$346$&$2.75$&$1.00$&$0.93 \pm 0.68$&$0.97 \pm 0.68$\\
$\texttt{BAM0126}\_\texttt{SLy}\_080\_275$&$365$&$2.75$&$0.80$&$1.50 \pm 1.21$&$1.18 \pm 0.82$\\
$\texttt{BAM0127}\_\texttt{SLy}\_066\_1275$&$407$&$2.75$&$0.66$&$0.95 \pm 0.70$&$0.96 \pm 0.71$\\
$\texttt{THC0063}\_\texttt{BLh}\_100\_273$&$511$&$2.73$&$1.00$&$0.85 \pm 0.57$&$0.92 \pm 0.61$\\
$\texttt{THC0064}\_\texttt{BLh}\_075\_277$&$430$&$2.77$&$0.75$&$0.88 \pm 0.59$&$0.94 \pm 0.63$\\
$\texttt{THC0038}\_\texttt{BLh}\_065\_280$&$312$&$2.80$&$0.65$&$0.83 \pm 0.53$&$0.81 \pm 0.48$\\
\br
\end{tabular}
\end{table}
We conclude that \texttt{SACRA} consistently achieves overall second or higher-order convergence for equal-mass and near-equal-mass binaries, while \texttt{BAM} and \texttt{THC} demonstrate first-order or even lower convergence, regardless of the mass ratio.

\subsection{Universal Relations and Analytical Fits}

\subsubsection{Identification of Key Frequencies}

While NR codes simulating BNS mergers are reliable and convergent during the inspiral, they tend to lose convergence in the post-merger phase. To better quantify code performance in this regime, we focus on evaluating how accurately the simulated waveforms predict QURs. 
Our initial task is to define a systematic approach for identifying the three key frequencies: $f_{mrg}$, $f_{max}$ and $f_2$. 
For example, the merger frequency is found at the time of maximum amplitude for strain; however, for $\Psi_4$, this is determined by the first peak in amplitude as discussed in \cite{arXiv:2310.10728}, and requires a slightly modified identification process. For strain, the merger time can be efficiently determined numerically using a general maximum function applied to amplitude data. 
For $\Psi_4$, identifying the merger requires a peak-finding function with a threshold to filter out minor peaks arising from numerical noise.

Another challenge in identifying the key frequencies comes from the instabilities near the maximum frequency reached in the post-merger. 
The time at which the GW frequency reaches its maximum corresponds to the point when the amplitude reaches a sharp minimum. This rapid swing in amplitude -- from maximum to minimum -- occurs over an extremely short interval, and many codes struggle to resolve it accurately \cite{arXiv:1306.4065}. Consequently, large instabilities appear in this frequency, often yielding non-physical values, including negative spikes. This complicates the accurate identification of $f_{max}$ and to address it, we focus on the time around the first minimum in amplitude immediately following the merger. We use a peak-finding function with an iteratively decreasing threshold to locate the first significant peak in amplitude after the merger time, followed by identifying the minimum amplitude between this peak and the merger amplitude. This approach ensures that we capture the relevant minimum amplitude associated with the maximum frequency.

\begin{figure}[h]
    \centering
    \includegraphics[width=0.9\linewidth]{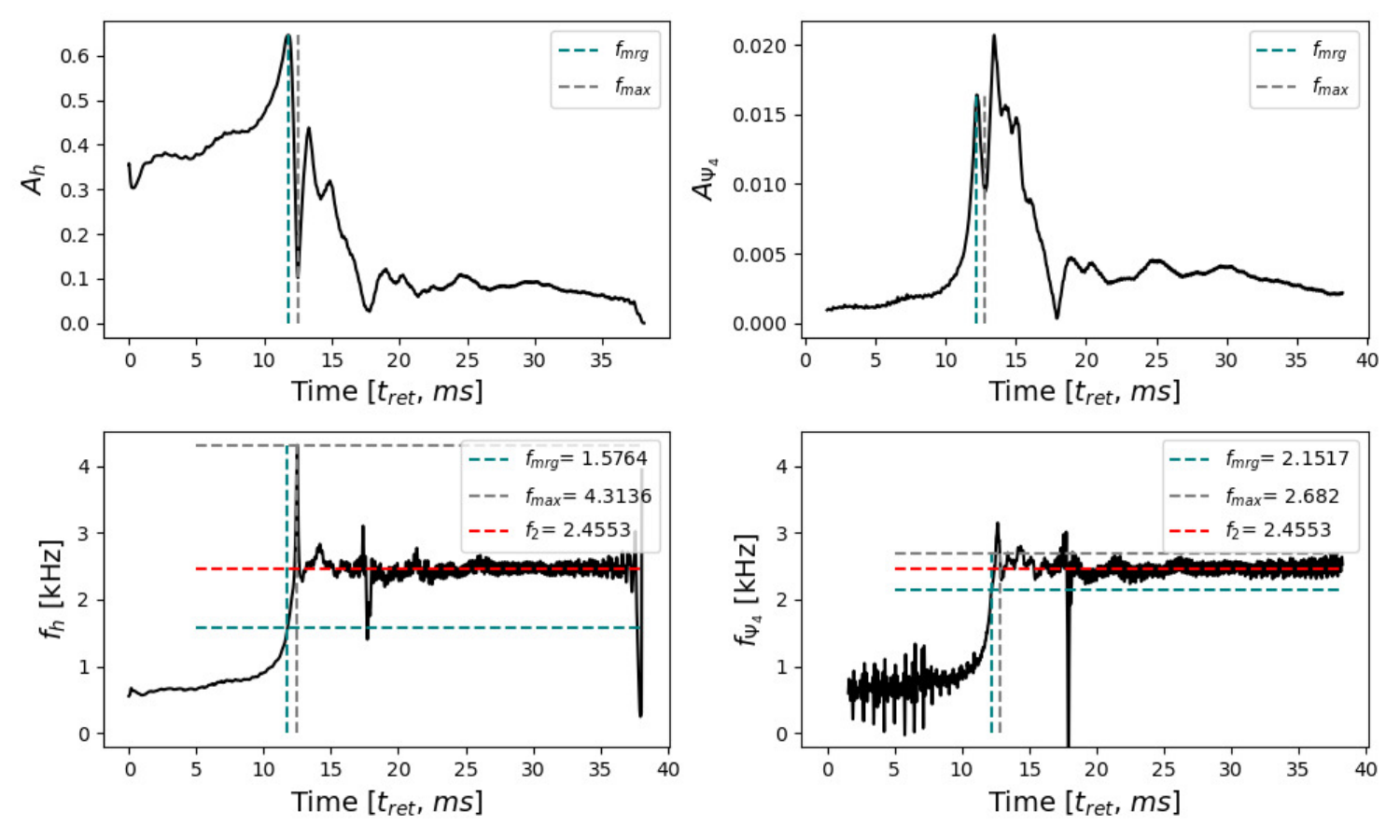}
    \caption{Example of frequency identification from amplitude plots for both strain and Weyl scalar in a sample BNS system (\texttt{THC0041\_DD2\_149\_125}).}
    \label{fig:figure5}
\end{figure}

\begin{figure}[h]
    \centering
    \includegraphics[width=0.9\linewidth]{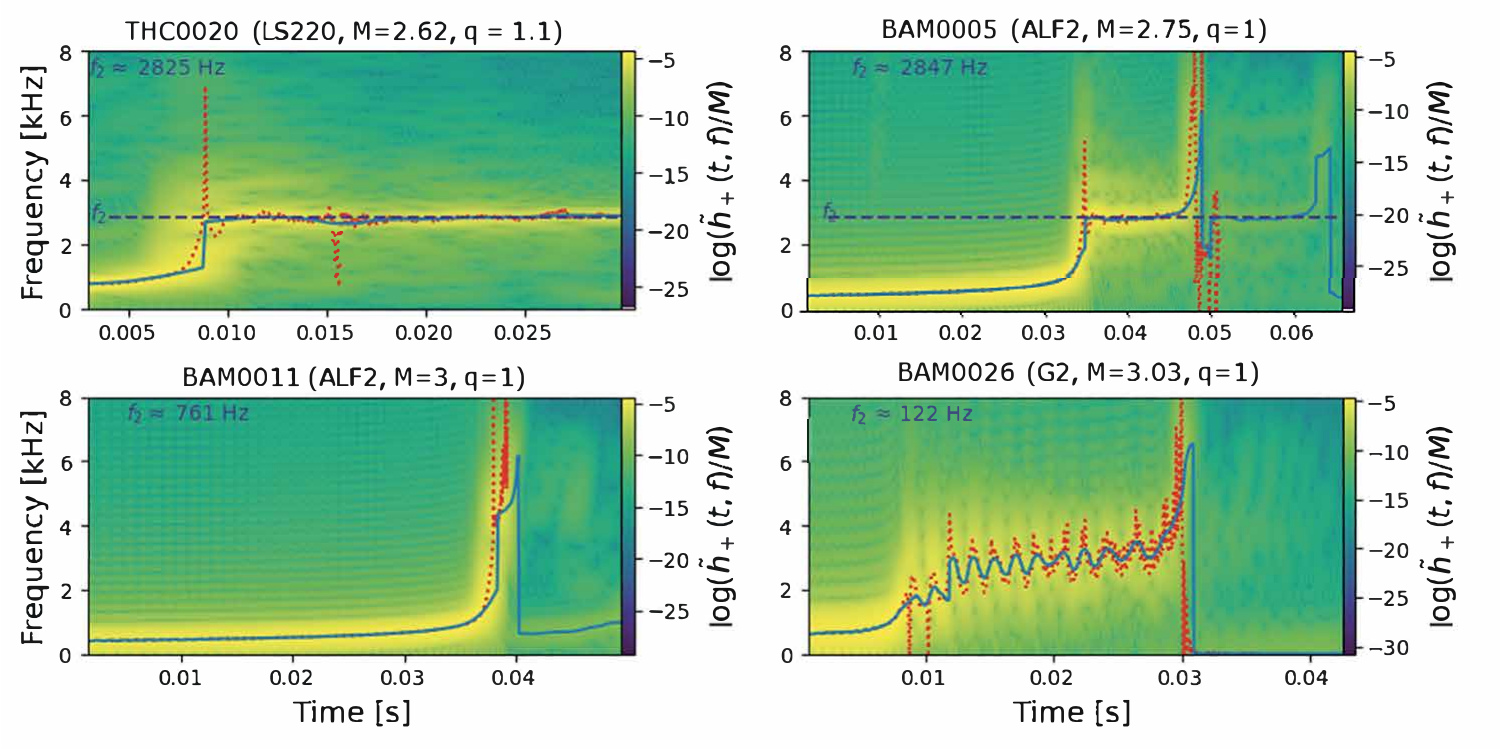}
    \caption{Spectrogram of sample strain waveforms, with instantaneous frequency shown in red and dominant frequency in each FFT window shown in blue.}
    \label{fig:figure6}
\end{figure}

We exemplify in Figure \ref{fig:figure5} the identification of the merger frequency $f_{mrg}$, and of the maximum frequency $f_{max}$ from the key features in the amplitude of the GW strain (upper left) and $\Psi_4$ (upper right).
While frequencies derived from the GW strain and the Newman-Penrose scalar $\Psi_4$ theoretically are the same physical quantity, numerical noise introduces an artificial distinction between them. 
In contrast to the $f_{mrg}$ and $f_{max}$ frequencies, who represent instantaneous frequencies tied to distinct moments in the merger process, the dominant frequency in the post-merger phase $f_2$, is typically a long-lived, monochromatic frequency. 
As seen in Figure \ref{fig:figure5} (lower plots),  $f_2$ can, in principle, be approximated from the instantaneous frequency,  however, this approach is generally unreliable due to noise in the post-merger region. 
Instead, to identify $f_2$, we generate a spectrogram for each system, providing PSD over time. This spectrogram is constructed by applying Fast Fourier Transforms (FFT) to a series of partially overlapping time intervals. We then identify $f_2$ by averaging the PSD across each frequency bin in the post-merger region and selecting the frequency at the peak of the averaged PSD values. Examples of this approach are illustrated in Figure \ref{fig:figure6}, which presents spectrograms for four BNS systems showing different behaviors. 

One challenge we faced was identifying $f_2$ in systems with extended post-merger data or high masses was when the remnant collapsed into a black hole.  
This is indicated by a sharp increase in frequency, followed by a rapid drop to zero as the remnant black hole stabilizes, 
as seen in three of the spectrograms from Figure \ref{fig:figure6}. Here, the dominant frequency band rises asymptotically before disappearing, leaving no dominant frequency in the spectrum. The lower left spectrogram in this figure reveals a prompt collapse to a black hole, which is expected because the total mass of the system is above the threshold mass. 
Both plots on the right depict short-lived post-merger remnants that eventually collapse. 
The spectrogram in the lower right shows an oscillatory frequency persisting through the post-merger phase until the collapse. We observe that while the system with this oscillatory post-merger frequency has a mass similar to the promptly collapsing system on the lower left, its tidal deformability is significantly higher, indicating a stiffer EOS.
This is consistent with previous studies, showing that the threshold mass for prompt collapse is higher for less compact neutron stars with greater tidal deformabilities  \cite{arXiv:1901.09977}. This large deformability supports a short-lived post-merger remnant and may also contribute to the oscillatory frequency observed, due to enhanced tidal effects.

Identifying $f_2$ was not possible for systems that promptly collapsed.
For systems with short-lived neutron star remnants, we began the spectrogram at the start of the simulation, which typically allowed for effective $f_2$ identification. 
However, some short-lived remnants had oscillatory or upward-sloping frequency curves, neither of which contained a monochromatic $f_2$ frequency. This behavior might suggest a phase transition that increases the dominant post-merger frequency.
Conversely, starting the spectrogram at the beginning of the data limited accurate $f_2$ identification for a subset of simulations with long pre-merger data.
Additional challenges arose in systems where the frequency spectrum was diffuse, due to the shredding of the lower-mass star in unequal-mass binaries at the end of the inspiral.

Due to the difficulty in reliably identifying a stable dominant frequency $f_2$ via the spectrogram for numerous systems, we plotted the instantaneous frequency on the spectrogram and used it to separate the systems with consistent, stable $f_2$ identification from those where $f_2$ was unidentifiable. This method allowed us to include only those systems with reliably identified $f_2$ in our analysis.
We note that all \texttt{SpEC} waveforms were excluded because they lacked sufficient post-merger data to observe a clear $f_2$ frequency. 
Since \texttt{SACRA} data ends shortly after the merger, we excluded most pre-merger data from the spectrogram to effectively identify $f_2$.

Extracting key frequencies from the $\Psi_4$ waveform was challenging due to inconsistent time scaling by the total mass in a subset of \texttt{BAM} data. 
To address this, we implemented a filter to identify differently scaled data, based on the expectation that $f_{h, 2} \approx f_{\Psi_4, 2}$, as shown in Figure \ref{fig:figure5} and detailed in \cite{arXiv:2310.10728}. This filter checks if $f_{\Psi_4, 2}$ is more than twice $f_{h, 2}$, given that all system have a total mass greater than $2 M_{\odot}$. 
However, this filter was not suitable for $f_{\Psi_4, mrg}$ and $f_{\Psi_4, max}$ because their values significant deviate from the corresponding strain values. 
Instead, we plotted these frequencies against tidal deformability, with and without \texttt{BAM} data, to identify outliers that deviate from the main trend. After correcting these frequencies, we continued with the analysis.

\subsection{Analytical Fits}

Using the identified frequencies, we assessed each code's alignment with the QURs that link frequencies to the EOS-dependent parameter, the tidal deformability, as explored in previous studies \cite{arXiv:1604.00246, arXiv:1907.03790, arXiv:2210.16366, arXiv:2310.10728}.
As previously discussed, we expect these frequencies to be dependent on the total mass as well. 
To accommodate this, we rescaled the dimensionless frequency in eq.(\ref{eq:QUR}), to ${\hat f_{key}} = \frac{f_{key} M}{\texttt{Hz }M_{\odot}}$, to account for scaling effects with the total mass, working with the rescaled, dimensionless form of the key frequencies:
\begin{equation}
(\hat f_{mrg}, \hat f_{max}, \hat f_{2}) = \frac{M}{\texttt{Hz}M_{\odot}}(f_{mrg}, f_{max}, f_{2}).
\end{equation}
For equal-mass binaries, we applied a QUR fit of the form:
\begin{equation}
\log_{10} \left ( \hat f_{mrg}, \hat f_{max}, \hat f_{2} \right ) = a + b \tilde \Lambda^{1/5}.
\end{equation}
For analyses that included unequal-mass binaries, we extended the fit to:
\begin{equation}
\log_{10} \left ( \hat f_{mrg}, \hat f_{max}, \hat f_{2} \right ) = a + (b_0 + b_1 q + b_2 q^2) \tilde \Lambda^{1/5}.
\end{equation}
Additionally, we compared these fits to those previously reported in \cite{arXiv:1604.00246, arXiv:1907.03790, arXiv:2310.10728}, using coefficients for the equal-mass binary case, as detailed in Table \ref{tab:tab4}.

\begin{table}[h]
\centering
\caption{\label{tab:tab4}Previously reported QUR functions and coefficients, for equal-mass binaries.}
\begin{tabular}{@{}l l l l l}
\br
Identifier & Formula & Coefficient & Coefficient & Frequency \\
\mr
{Fit 1}$^a$ & $a_0 + a_1 \tilde \Lambda^{1/5}$ & $a_0=4.242$ & $a_1=-0.154$ & $f_{h, mrg}$\\
{Fit 2}$^b$ & $a_0 + a_1(k_2^T)^{1/5}$ & $a_0=4.186$ & $a_1=-0.195$ & $f_{h, mrg}$\\
{Fit 3}$^c$ & $a_0(\eta) + a_1(\eta)\tilde \Lambda^{1/5}$ & $a_0=4.2285$ & $a_1=-0.149$ & $f_{h, mrg}$\\
{Fit 4}$^d$ & $a_0 +(b_0+b_1q+b_2q^2) (k_2^T)^{1/5}$ & $a_0=4.589$ & $\sum b_{i |(q=1)}=-0.274$ & $f_{\Psi_4, mrg}$\\
{Fit 5}$^e$ & $a_0 + (b_0+b_1q+b_2q^2) (k_2^T)^{0.06}$ & $a_0=6.550$ & $\sum b_{i |(q=1)}=-1.885$ & $f_{\Psi_4, max}$\\
{Fit 6}$^g$ & $a_0 + (b_0+b_1q+b_2q^2) (k_2^T)^{1/5}$ & $a_0=4.617$ & $\sum b_{i |(q=1)}=-0.274$ & $f_{h, 2};~f_{\Psi_4, 2}$\\
{Fit 7}$^f$& $c_0(\eta) + c_1(\eta)\tilde \Lambda^{1/5}$ & $c_0=4.5085$ & $c_1=-0.17275$ & $f_{h, 2}$\\
\br
\end{tabular}\\
$^{a}$ \cite{arXiv:1412.3240};
$^{b, d}$ $k_2^T = (3/16)\tilde \Lambda$ \cite{arXiv:1604.00246};
$^{c, f}$ $\eta = 0.25$ \cite{arXiv:1907.03790};
$^{e, g, h}$ $k_2^T = (3/16)\tilde \Lambda$ \cite{arXiv:2310.10728}.
\end{table}
Note that in the figures, $\Lambda$ corresponds to $\tilde \Lambda$ as used in the text.

\subsubsection{QUR fits for Merger Frequency}

\begin{figure}[h]
   \centering
    \includegraphics[width=0.9\linewidth]{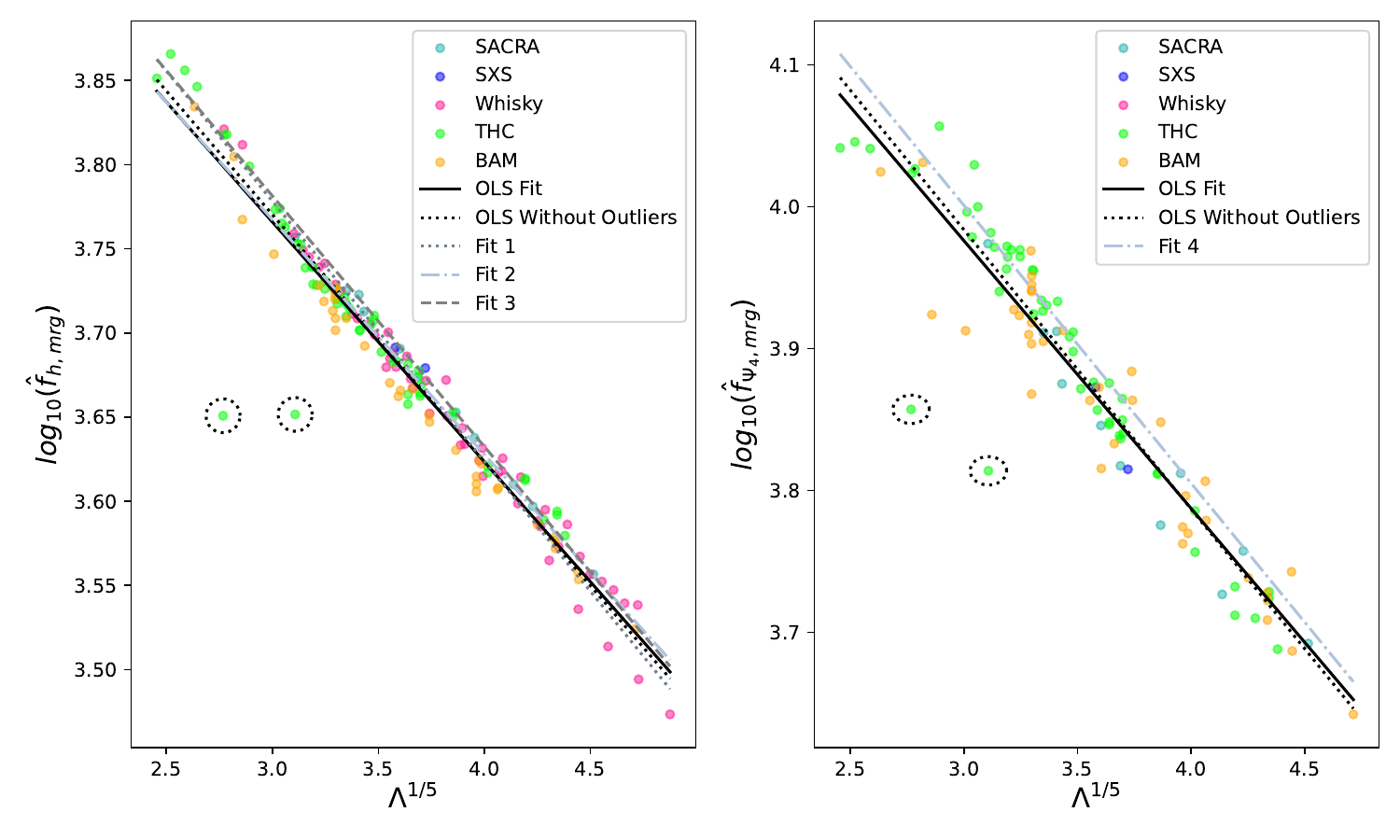} 
    \caption{QUR fits for $f_{mrg}$ with equal-mass binaries ($q=1$) using strain data (left) and for $\Psi_4$ data (right). Each plot includes two OLS fits: one with all data points and another excluding circled outliers. Previous fits are included for comparison.}
    \label{fig:figure7}
\end{figure}

\begin{figure}[h]
    \centering
    \includegraphics[width=0.90\linewidth]{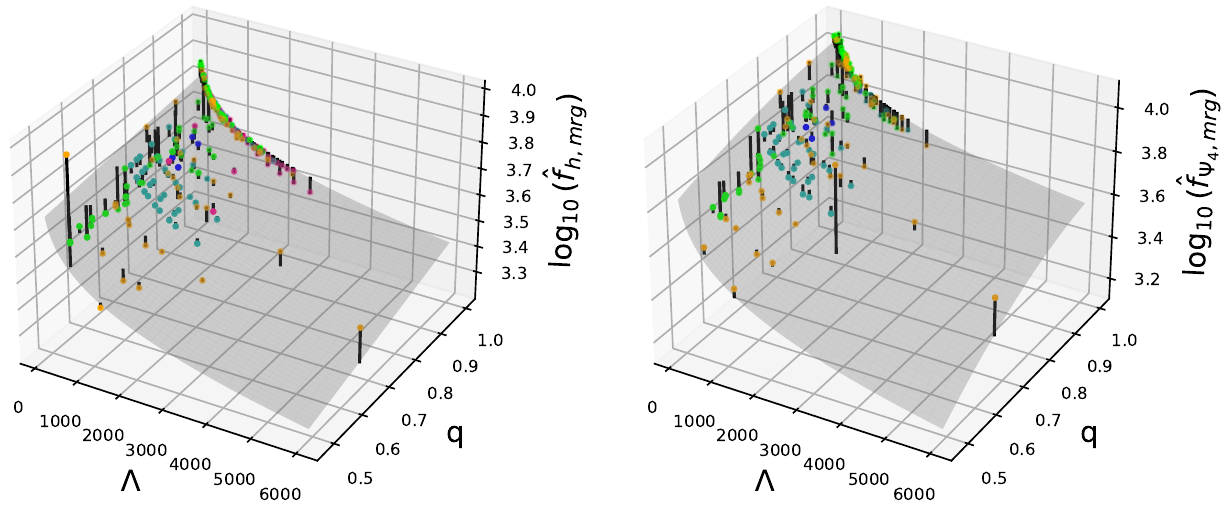}
    \caption{Mass-ratio dependent QUR fits for strain data (left) and $\Psi_4$ data (right), including both equal-mass and unequal-mass binaries.}
    \label{fig:figure8}
\end{figure}

We begin our analysis with the merger frequency, $f_{mrg}$, which corresponds to the maximum amplitude in the strain. 
Figure \ref{fig:figure7} presents scatter plots comparing our fits for the QUR that relates $f_{mrg}$ to the EOS through tidal deformability, alongside previously reported fits for equal-mass binaries from Fit 1 \cite{arXiv:1412.3240}, Fit 2 \cite{arXiv:1604.00246}, Fit 3 \cite{arXiv:1907.03790} and Fit 4 \cite{arXiv:2310.10728}. 
The left panel of Figure \ref{fig:figure7} (left) displays the QUR fits to $f_{mrg}$ derived from strain data, while the right panel shows the fits for the $\Psi_4$ data.
For each dataset, we conducted two ordinary least squares (OLS) analyses: one including all data points and another excluding two outliers identified in the \texttt{THC} data, which are circled in the plot.
We the extend our fits to unequal-mass binaries, shown as surface plots in Figure \ref{fig:figure8} for strain (left) and $\Psi_4$ (right).

Table \ref{tab:tabA1} in \ref{sec:appendix} presents the coefficients of our fits to $f_{mrg}$ for the equal-mass binaries, while Table \ref{tab:tabA2} extends these fits to unequal-mass binaries. 
These tables also include the coefficient of determination ($R^2$) for each fit, defined as:
\begin{equation}
R^2 = 1 - \frac{\sum (f^{num}-f^{fit})^2}{\sum(f^{num} -\bar{f}^{num})^2}\textcolor{blue}{\footnotemark}
\end{equation} 
\footnotetext{See \href{https://www.statsmodels.org/stable/index.html}{statsmodels}}
The results from Table \ref{tab:tabA1} indicate that for equal-mass binaries, our $f_{h, mrg}$ fit demonstrates a strong correlation ($R^2=0.944$), closely aligning with previous findings. 

After removing the marked outliers, the fit slightly improves, achieving an $R^2$ of $0.979$ and aligning well with the central trend of the other fits.
The correlations for $f_{\Psi_4,mrg}$ are slightly lower, with $R^2$ values of $0.884$ and $0.934$ with and without outliers, respectively, yet they still indicate strong correlations.
Our fit for the $q$-dependent case, as shown in Table \ref{tab:tabA2}, reveals that including unequal-mass binaries results in a weaker correlation ($R^2=0.753$ for strain and $R^2=0.758$ for $\Psi_4$ data). Notably, a significant portion of the fitting error arises from systems with low tidal deformability, although several such systems still align closely with the fit surface. The most substantial deviations from the quasi-universal relation fits are observed in the \texttt{THC} and \texttt{BAM} datasets, which include the points with the greatest discrepancies.

We assessed each code’s performance individually and against a collective fit for $f_{mrg}$, presenting the joint fitting coefficients for strain in Table \ref{tab:tabA3} and in Table \ref{tab:tabA4} for $\psi_4$. 
Additionally, Table \ref{tab:tabA3} includes the associated percent errors, with $\%$ errors relative to the fit obtained for equal-mass binaries (excluding outliers) provided in parentheses. In the equal mass case, the $b_i$ terms are combined into one term $b=\sum b_{i |(q=1)}$.
Notably, the highest errors originate from the \texttt{THC} dataset, followed by \texttt{BAM}, with all strain data -- except for \texttt{THC} -- showing less than $5\%$ error for the $q=1$ case. 
Despite considerable variability in the fits among q-dependent terms, particularly pronounced in the \texttt{Whisky} data, the reduction to the equal-mass case aligns well across codes, matching our combined fits for equal-mass binaries. This alignment suggests that the tidal deformability dependence is well captured in the equal-mass scenario. Only \texttt{THC} data significantly deviates from these trends, indicating less consistency with QURs, especially in the unequal-mass case.

\subsubsection{QUR fits for Maximum Frequency}
The maximum frequency, $f_{max}$, presents significant challenges for accurate capture due to its occurrence near the amplitude minimum where shock-related instabilities are common, as noted in \cite{arXiv:1306.4065, arXiv:2310.10728}. Around this point, the instantaneous frequency often becomes ill-defined, displaying spurious spikes and troughs that lead to substantial frequency fluctuations.

Table \ref{tab:tabA1} shows that the fits for $f_{max}$ derived from strain data are imprecise, failing to capture the variability associated with tidal deformability, as reflected by low $R^2$ values of $0.252$ and $0.131$. 
Shock-related instabilities likely account for the observed variations. These instabilities can introduce numerical artifacts and irregularities in the extracted waveform, leading to variations in the fitted frequency.
However, we note that up to $\Lambda^{1/5} \approx 4$, $f_{max}$ appears approximately constant, suggesting minimal or no dependence on tidal deformability, averaging between $Mf_{max} \approx 10^{4}$ to $10^{4.5}$.
Given the high variability in maximum frequencies, establishing a clear dependence on $\Lambda$ is challenging, complicating the use of $f_{max}$ from strain data for code comparisons. Notably, \texttt{BAM} and \texttt{Whisky} data clusters significantly below others might indicate greater numerical instability in these codes around the amplitude minimum. 
Alternatively, this pattern could be attributed to specific physical properties of the EOS at high tidal deformabilities. The variance complicates drawing definitive conclusions.

\begin{figure}[h]
    \centering
    \includegraphics[width=0.90\linewidth]{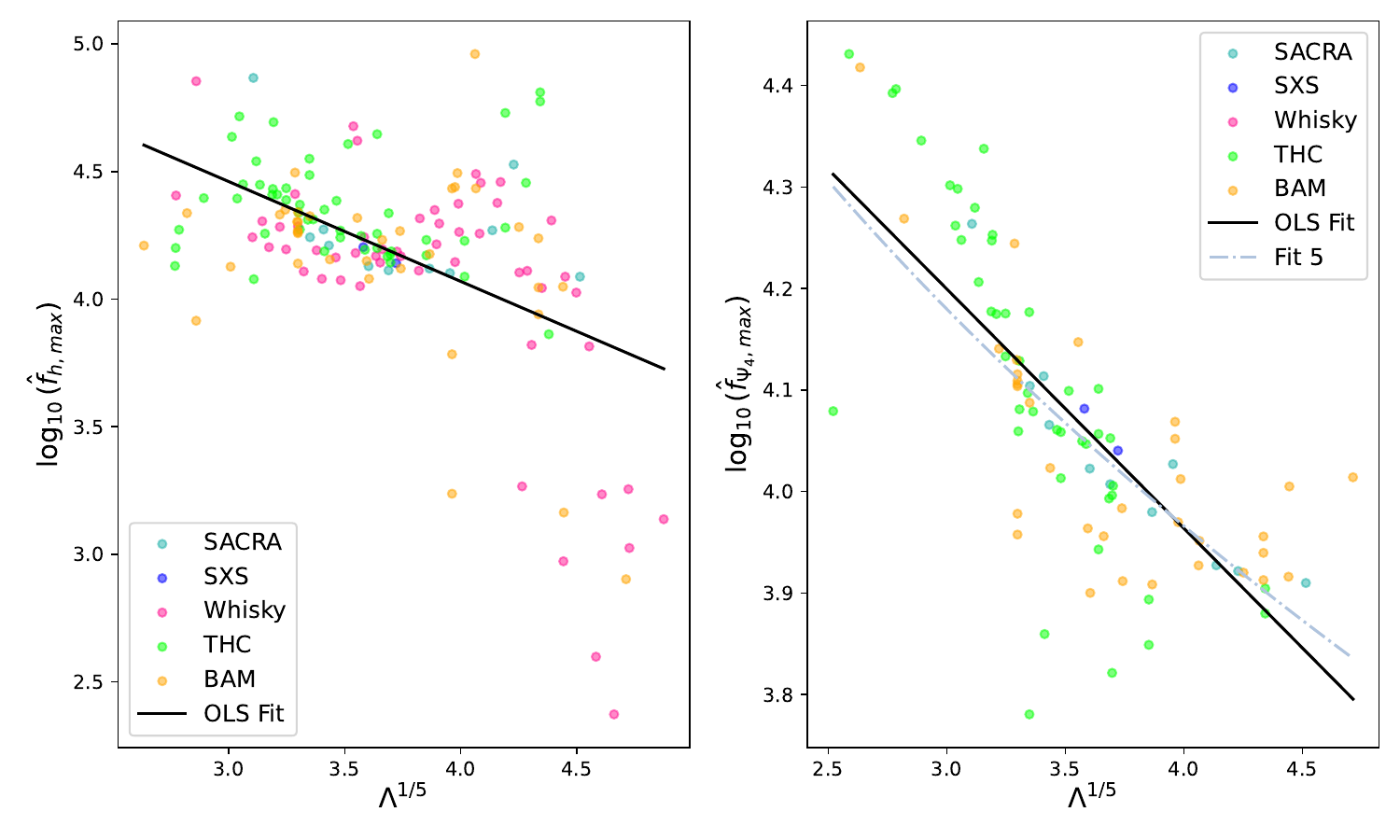}
    \caption{Scatter plots of $f_{max}$ for equal-mass binaries, testing the $\tilde \Lambda^{1/5}$ dependence in QUR. The left plot shows strain data, and in the right is $\Psi_4$ data. Our OLS fits are shown, with the fit reported in \cite{arXiv:2310.10728} included for $\Psi_4$ data.}
    \label{fig:figure9}
\end{figure}

Figures \ref{fig:figure9} presents scatter plots of our QUR fits for $f_{max}$ using equal-mass only data for both strain and $\Psi_4$, with the right panel including previously reported fits for $f_{\Psi_4,max}$ \cite{arXiv:2310.10728}. 
Table \ref{tab:tabA5} lists the fit parameters derived for the $f_{\Psi_4, max}$ QUR, accounting for mass ratio dependence for the individual codes.

\begin{figure}[h]
    \centering
    \includegraphics[width=0.90\linewidth]{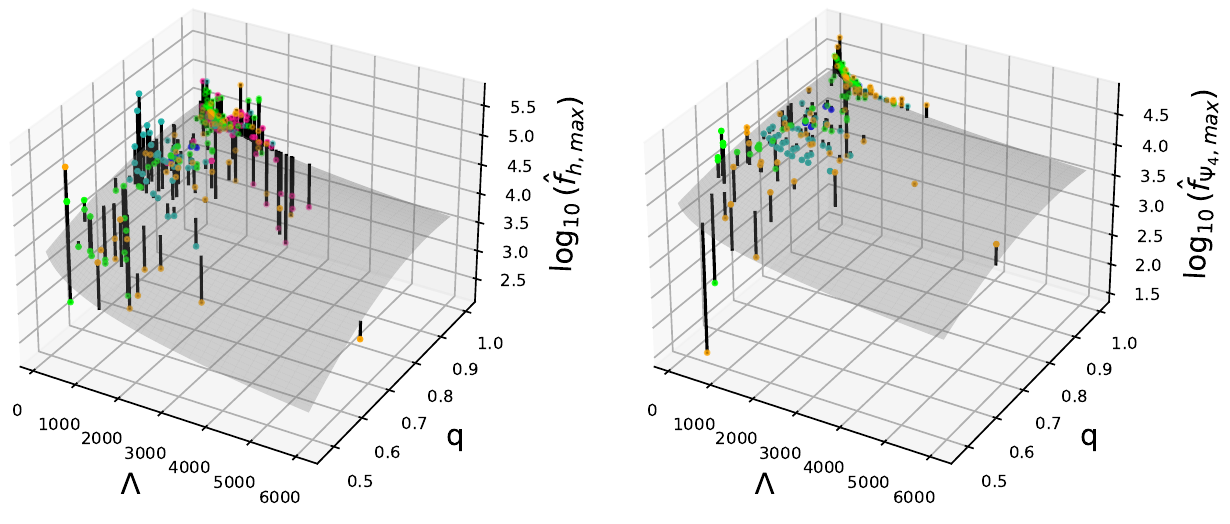}
    \caption{Surface plots testing the mass ratio-dependent QUR for $f_{max}$ data. Strain data is shown on the left, and $\Psi_4$ data on the right.}
    \label{fig:figure10}
\end{figure}

Figure \ref{fig:figure10} contains surface plots of our fits to $f_{max}$ data for unequal-mass binaries, using both strain and $\Psi_4$ data, showing the dependence on mass-ratio across codes.

\subsubsection{QUR fits for the Dominant Frequency}

We now turn to examining the dominant post-merger frequency, $f_2$, and the corresponding QUR fits. In the post-merger phase, this dominant emission frequency typically ranges from $2$ to $5$ kHz, positioned between $f_{mrg}$ and $f_{max}$. To accurately compare frequencies across simulations, we generate spectrograms covering both the inspiral (pre-merger) and post-merger phases.
This approach allows us to identify dominant frequencies and track their evolution over time.

\begin{figure}[h]
    \centering
    \includegraphics[width=0.90\linewidth]{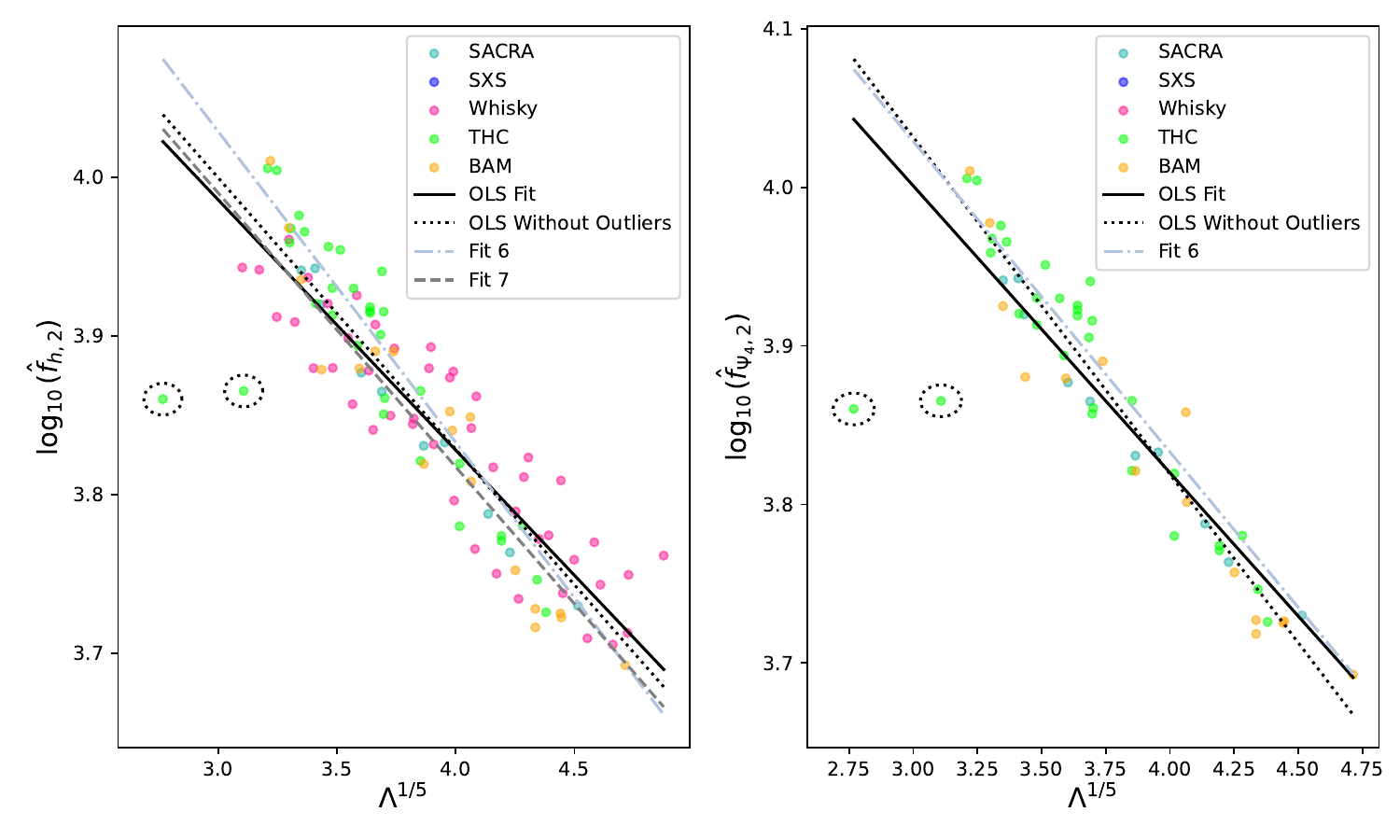}
    \caption{Scatter plots of $f_{2}$ for equal-mass binaries. Left plot: strain data; right plot: $\Psi_4$ data. Our two OLS fits, with and without marked outliers, are displayed alongside three previously reported fits for strain data and one for $\Psi_4$ data.}
    \label{fig:figure11}
\end{figure}

\begin{figure}[h]
    \centering
    \includegraphics[width=0.90\linewidth]{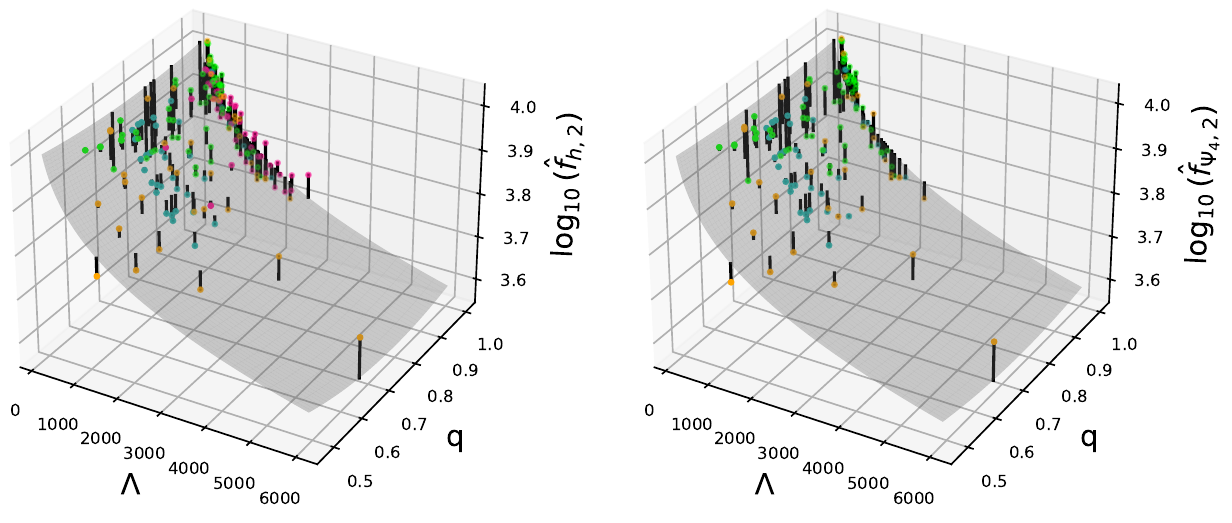}
    \caption{Left panel: surface fit for strain data including unequal-mass binaries. Right panel: mass-ratio dependent fit for $\psi_4$, also including unequal-mass binaries.}
    \label{fig:figure12}
\end{figure}

Figure \ref{fig:figure11} displays $f_2$ fits from both strain and $\Psi_4$ data for equal-mass binaries. Although the fit demonstrates good overall agreement, the data points show slightly more spread than those for the $f_{mrg}$ fit, likely due to increased numerical error and loss of convergence in the post-merger phase. This trend is expected, as most codes exhibit more reliable convergence before and during merger -- prior to the shocks at the amplitude minimum -- than in the post-merger phase \cite{arXiv:2307.03250}. 
For comparison, alongside our fit for equal-mass binaries, we plot in Figure \ref{fig:figure11} previously reported fits \cite{arXiv:2310.10728, arXiv:1907.03790}, with strain data on the left and $\Psi_4$ data on the right. 
Our fits align reasonably well with prior results, though the correlation is somewhat weaker than that for $f_{mrg}$, further highlighting the increased numerical error in the post-merger. 
A similar pattern is visible in the fits for unequal-mass binaries in Figure \ref{fig:figure12}, where outliers are more pronounced and the overall variability in the $f_2$ fit is greater compared to the $f_{mrg}$ fit. 
Interesting, the presence of a few significant outliers in the $q$-dependent $f_{mrg}$ fit reduces the $R^2$ to approximately the same value as the $f_2$ fit, as detailed in Table \ref{tab:tabA2}.
Table \ref{tab:tabA2} also shows considerable variability in the $q$-dependent terms, while the constant term and $\Lambda^{1/5}$ coefficient display less variation in the equal-mass limit. This likely reflects a data bias towards equal-mass binaries ($q=1$) compared to unequal-mass cases. A more balanced distribution across the $q$ parameter space could better constrain the $q$-dependent terms, potentially enhancing the characterization of code performance across different mass ratios. However, with the current dataset, distinguishing whether the variability in $q$-dependent terms is due to numerical errors or the distribution of mass ratios remains challenging.

In Table \ref{tab:tabA6}, we present the fit parameters for the $f_{h,2}$ QUR from strain data, including mass-ratio dependence for each  code. We also provide a separate fit that excludes \texttt{THC} due to its lower self-consistency with the QURs. Percentage errors are reported relative to the best $q$-dependent and equal-mass fits.
Table \ref{tab:tabA7} lists the fit parameters for the $f_{\Psi_4,2}$ QUR derived from $\Psi_4$ data, excluding \texttt{Whisky} due to its lack of $\Psi_4$ data and \texttt{SpEC}, which only provides data shortly beyond the merger.

We find that the best individual fits for $f_2$ came from \texttt{SACRA}, followed by \texttt{BAM} and then \texttt{Whisky}. This ranking contrasts with the $f_{mrg}$ fits, where \texttt{Whisky} was nearly as self-consistent as \texttt{SACRA}, suggesting that \texttt{Whisky}’s accuracy may be higher at the time of merger and lower in the post-merger phase compared to \texttt{BAM}. 
\texttt{THC} showed the least conformity to the QUR fits, with an $R^2$ of $0.315$,  indicating potential numerical errors or instability in this code during the post-merger phase. This discrepancy may also reflect differences in the physics captured by \texttt{THC} or limitations of the QUR for this dataset. 

\begin{figure}[h]
    \centering
    \includegraphics[width=0.90\linewidth]{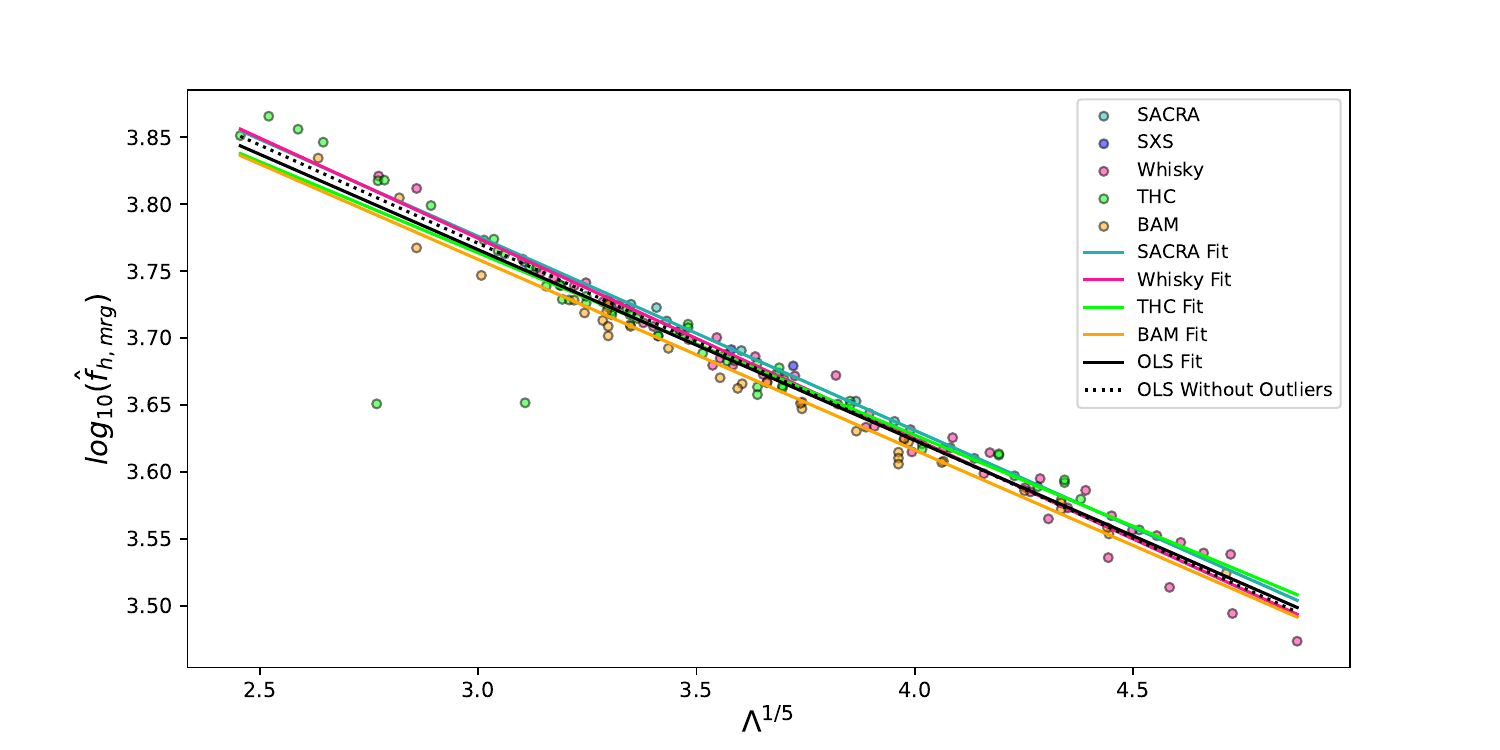}
    \caption{Scatter plot of $f_{mrg}$ for equal-mass binaries from strain data, with individual fits for each code. \texttt{SpEC} is excluded due to insufficient data.}
    \label{fig:figure13}
\end{figure}

As a summary test of the code performance relative to the QUR, Figure \ref{fig:figure13} displays individual fits of the $f_{h,mrg}$ relation for equal mass binaries for each code. This reveals close agreement among all codes, suggesting stable and consistent performance around the time of merger. 
The agreement between different methods in producing a QUR for $f_{h, mrg}$ supports the idea that this relation reflects a physical property of the system, rather than being an artifact of the computational models.

\subsubsection{QUR fits for Post-merger Transient Time}
In addition to examining the previously reported QURs for frequency \cite{arXiv:1604.00246, arXiv:1907.03790, arXiv:2210.16366, arXiv:2310.10728}, we have identified and analyzed for the first time a new QUR that links the time between the merger and the maximum frequency $\Delta t_{f_{max},f_{mrg}} = t_{f_{max}}-t_{f_{mrg}}$ with the dimensionless tidal deformability $\tilde \Lambda$. 
We choose to measure the time difference between these two events due to the wide range of initial conditions in our data, which could make the identification of only the merger time inconsistent without a reliable reference point for the initial time. 
The merger time, however, is a significant and easily distinguishable event in the waveform, allowing us to standardize it to $t_{mrg} = 0$. The time difference between the merger and maximum frequency can then be simply expressed as $t_{f_{max}}$. 
Based on the results of our analysis, we propose an additional QUR for the time of maximum frequency relative to the merger time, in the form:
\begin{equation}
\label{tmax_QUR}
    \log_{10} \left ( \hat {t}_{f_{max}} \right ) = \log_{10} \left ( t_{f_{max}}/M \right ) = a_0 + (b_0+b_1q+b_2q^2)(\tilde \Lambda)^{1/5}.
\end{equation}
 
\begin{figure}[h]
    \centering
    \includegraphics[width=0.90\linewidth]{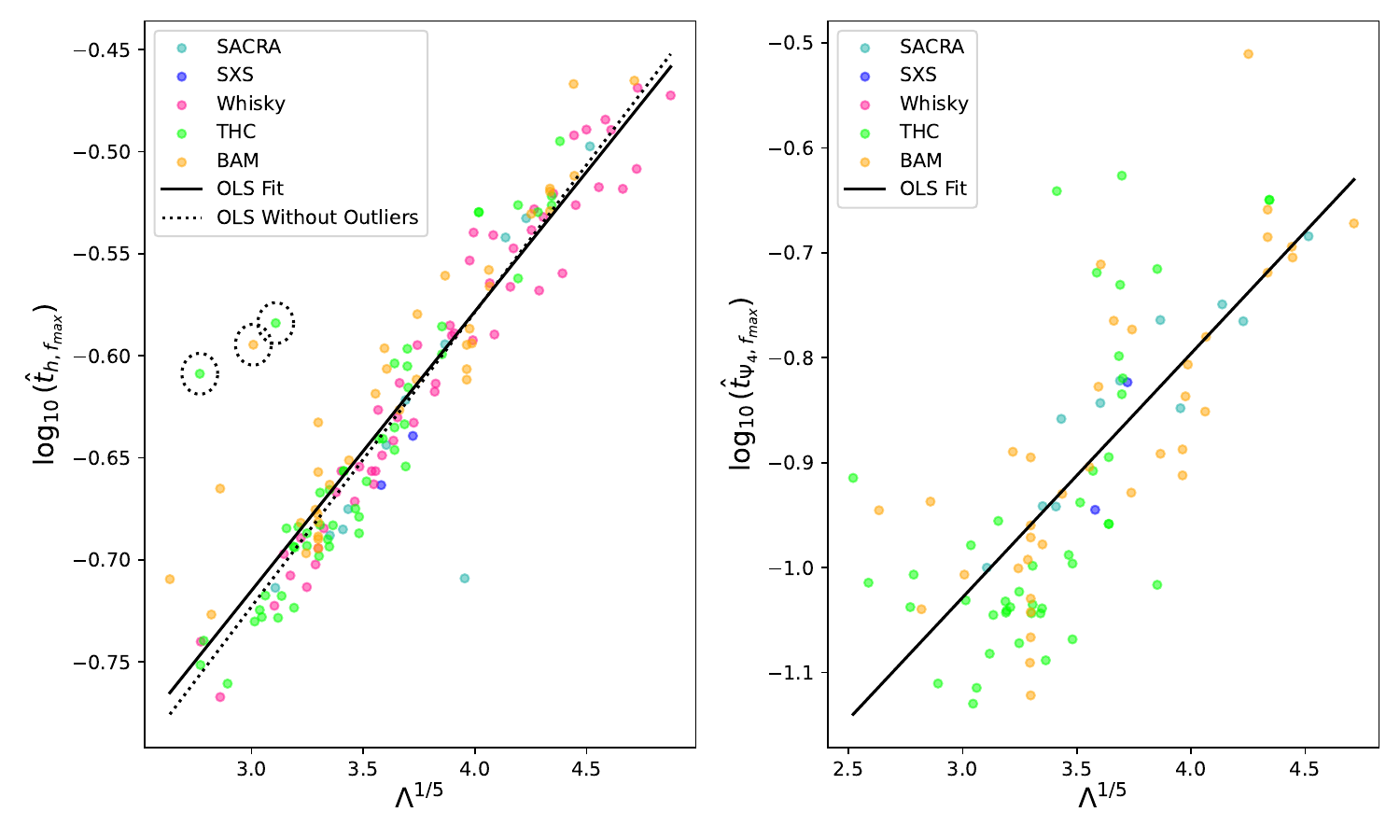}
    \caption{Scatter plots of $t_{f_{max}}$ for equal-mass binaries. Left plot: strain data with two OLS fits, with and without outliers. Right plot: $\Psi_4$ data with one OLS fit.}
    \label{fig:figure14}
\end{figure}

\begin{figure}[h]
    \centering
    \includegraphics[width=0.90\linewidth]{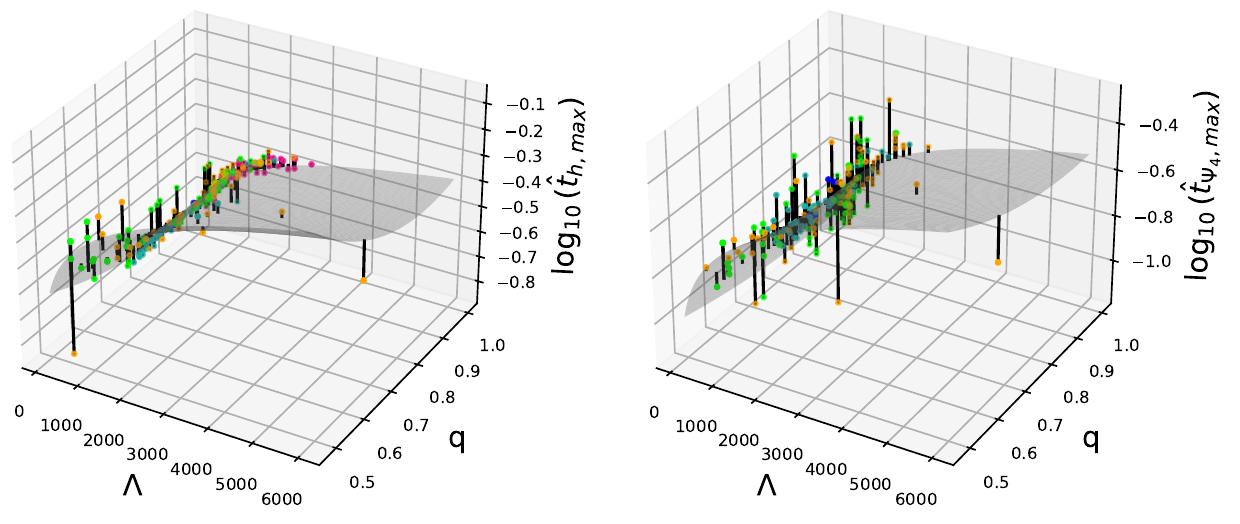}
    \caption{Surface fit for $t_{f_{max}}$, with data including unequal-mass binaries. Left plot: $q$-dependent QUR for strain. Right plot: $q$-dependent QUR for $\psi_4$.}
    \label{fig:figure15}
\end{figure}

In Figure \ref{fig:figure14} we plot this transient time $t_{f_{max}}$ against the effective tidal deformability $\tilde \Lambda$ using the subset of equal mass binaries for both strain (left) and $\Psi_4$ (right), including OLS fits with and without outliers. We extend these fits to include unequal mass systems in Figure \ref{fig:figure15}.
From Figure \ref{fig:figure14}, it is evident that the time of this transient region increases with tidal deformability, contrasting with the behavior of frequencies. This relationship shows how the timescale of this process is influenced by the effect of tidal interactions affecting the dynamics immediately post-merger. This is relevant for understanding how the evolution of the remnant -- from the maximum GW amplitude at the merger to the closest approach of the cores at the minimum amplitude -- is dependent on the EOS. 

\begin{figure}[h]
    \centering
    \includegraphics[width=0.90\linewidth]{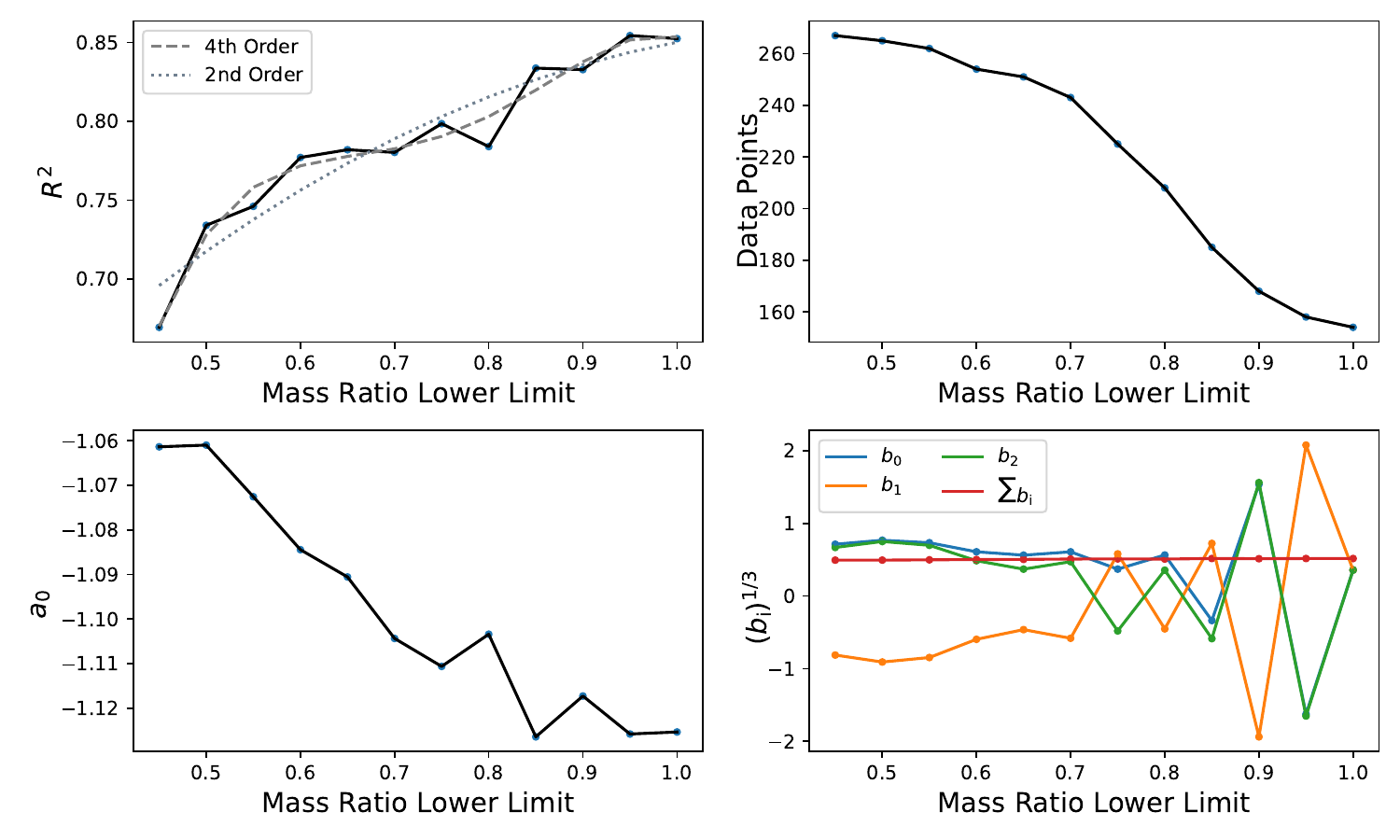}
    \caption{
Analysis of QUR fits for $t_{f_{h,max}}$ by varying the mass ratio lower limit. Top Left: $R^2$ values of the fits vs. mass ratio. Top Right: Data point count vs. mass ratio. Bottom Left: $a_0$ term vs. mass ratio, plotted on a cube root scale. Bottom Right: $b_n$ coefficients vs mass ratio on a cube root scale to distinguish small variations.}
    \label{fig:figure16}
\end{figure}

We present the QUR fits for $t_{f_{max}}$ for equal-mass binaries for both strain and $\Psi_4$ data in Table \ref{tab:tabA1}, and the mass ratio-dependent fits in Table \ref{tab:tabA2}. 
The QUR for equal mass binaries from strain data shows strong correlations, with $R^2$ values of $0.852$ and $0.911$, with and without outliers, respectively. 
However, including unequal mass binaries, the correlation for the mass ratio-dependent fit decreases to $R^2=0.669$. 
Yet, as shown in the top right plot of Figure \ref{fig:figure16}, removing the lower, more uneven mass ratios increases the correlation $R^2$, suggesting that the proposed QUR begins to break down at more extreme mass ratios. 
Figure \ref{fig:figure16} also indicates that, with the inclusion of a broad range of mass ratios, the $q$-dependent terms in eq.(\ref{tmax_QUR}) remain consistent. 
However, the variability in the $b_i$ coefficients increases when data contains only $q>0.7$ binaries.
This could be due to a high concentration of equal-mass points in our data set or to numerical instabilities in the codes. 
In Table \ref{tab:tabA9}, we provide fits and $R^2$ values for this mass ratio-dependent QURs for individual codes.
\texttt{SACRA}, \texttt{Whisky} and their combined fit with \texttt{SpEC} (which lacked sufficient data for an individual fit) all achieved $R^2>0.9$.
In contrast, \texttt{BAM} and \texttt{THC} yielded weaker correlations of $0.664$ and $0.498$ respectively.
This discrepancy is likely because \texttt{THC} and \texttt{BAM} codes contain lower mass ratios, where the QUR degrades. 

Extending the QUR fits to $\Psi_4$ has proven less effective, yielding only modest correlations around $R^2=0.5$ for the overall as well as \texttt{SACRA} and \texttt{BAM} fits, as shown in Table \ref{tab:tabA9}. The \texttt{THC} fit is notably lower, with an $R^2$ of less than $0.25$. This lower performance is likely caused by numerical errors in the $\Psi_4$ data within the transient region, which complicates the accurate identification of the maximum frequency compared to strain data.

We observe that the time QUR for strain, $t_{h,f_{max}}$, maintains high correlations for datapoints from moderate mass ratio to equal mass ratio.
As result, we propose our best-fit model for equal mass binaries with outliers removed as follows:
\begin{equation}
    \log_{10} \left ( \hat{t} _{h,f_{max}} \right ) = -1.1557+0.1443 \left (\tilde \Lambda \right )^{1/5},
\end{equation}
where $\hat{t}=t/M$.
When including unequal-mass binaries, we propose the following best fit for \texttt{SACRA}, \texttt{Whisky} and \texttt{SpEC}, which includes mass ratio $q\ge 0.73$.
\begin{equation}
    \log_{10} \left ( \hat{t} _{h,f_{max}} \right ) = -1.2098+\left (0.1907-0.0670q+0.0331q^2 \right ) \left (\tilde \Lambda \right )^{1/5}
\end{equation}
Data points from simulations using \texttt{BAM} and \texttt{THC} codes were excluded from this fit, due to their higher masses and more extreme mass ratios.

These models are particularly valuable as they provide information about the remnant dynamics shortly after the merger. 
In this phase, the sharp decrease in gravitational wave amplitude leads to instabilities in frequency data, marked by abrupt, unphysical spikes. The discrepancy in $f_{max}$ contrasts with the strong QUR observed for time in the same region. 
Thus, this time-based measure likely offers more reliable insights despite the numerical instabilities that complicate frequency-based inferences. This time-based relation provides a robust framework for understanding this part of the waveform, despite the numerical challenges associated with frequency calculations.

\section{\label{sec:conclusion}Conclusions}

We conducted a study of open GW data produced by five major NR codes that simulate BNS mergers:  \texttt{SACRA}, \texttt{BAM}, \texttt{THC}, \texttt{Whisky}, and \texttt{SpEC}, under the assumption that extraction methods and numerical techniques do not introduce major biases in the results. 
\textcolor{blue}{This is supported by previous studies showing that truncation error due to numerical discretization, particularly near merger, are the dominant sources of error \cite{arXiv:1301.3555, arXiv:1604.07999, arXiv:1803.07965}}.
Our analysis examined the convergence of the data, and assessed the accuracy of the simulations by quantifying numerical discrepancies in the dependence of the waveform parameters on tidal deformability.
Through systematic comparisons of independent simulations, we quantified systematic errors and assessed the consistency of results across using $R^2$. In the post-merger phase, differences in the QURs for key frequencies indicate variations between codes that could impact the construction of high-accuracy GW templates for tidal deformability extraction in future observations.

Typically, numerical codes report attaining up to second-order convergence during the inspiral phase and first-order or consistent results post-merger.
Our convergence analysis across the entire evolutionary domain of BNS mergers employed both monotonic and oscillatory convergence methods. \texttt{SACRA} achieved second-order or higher convergence, while \texttt{BAM} and \texttt{THC} generally exhibited first-order convergence. This varied convergence highlights the different capabilities of the codes in handling the complex dynamics of BNS mergers, particularly in the post-merger phase.
To improve simulation accuracy and code performance, achieving better convergence in the post-merger regime remains a priority. This could be addressed through higher-resolution simulations, improved wave extraction techniques, and refined numerical treatments of matter effects.

Overall, we achieved strong fits for $f_{mrg}$ and $f_2$ in equal-mass scenarios, and reasonable fits for unequal-mass cases.
However, the fits for $f_{max}$ were poor, including those for $\Psi_4$ data. 
The slightly weaker fits for $f_2$ in compared to $f_{mrg}$ indicate a common loss of convergence in the post-merger phase across most codes, although \texttt{SACRA} appeared less affected, closely followed by \texttt{Whisky} for $f_{mrg}$ fits and \texttt{BAM} for $f_2$ fits. 
Conversely, \texttt{THC} had greater numerical errors in the post-merger, indicating potential areas for improvement.
We had insufficient data for \texttt{SpEC} to draw definitive conclusions.

We introduced a new time-based relation for the transient time immediately following the merger, complementing traditional QURs on frequency. This provides valuable insights into remnant dynamics shortly after the merger, contrasting with $f_{max}$, which did not demonstrate a clear QUR, likely due to numerical instabilities. 
We have developed specialized Python scripts for data processing and analysis in the Jupyter environment, available as open source on our GitHub repository (\href{https://github.com/mbabiuc/CodeComp}{https://github.com/mbabiuc/CodeComp}). 

Looking forward, we plan to refine our convergence assessments, separate pre- and post-merger analyses, and expand our selection of EOS and mass ratios. 
We will also quantify the effects of residual eccentricity and the impacts of magnetic fields and neutrinos. These efforts aim to deepen our knowledge of code performance across different stages of BNS evolution, especially in the post-merger.
Our work advances our understanding of BNS mergers and sets the stage for future improvements in NR code development. These improvements will enhance waveform accuracy and reduce systematic errors in measuring tidal effects in GW observations, thus preparing codes for the accuracy demands on simulations imposed by the next generation GW detectors.

\ack

This research was made possible in part by the NSF REU Site: Appalachian Mathematics and Physics Award No. 2349289 to Marshall University.
MCBH also acknowledges support from the NASA Established Program to Stimulate Competitive Research, Grant No. 80NSSC22M0027, and from the NSF grant No. PHY-1748958 to the Kavli Institute for Theoretical Physics.


\appendix
\section{\label{sec:appendix}Quasi-universal Relations Fits}

\begin{table}[h]
\centering
\caption{\label{tab:tabA1}Summary of fit parameters to QUR for key frequencies and transient times across all codes for equal-mass binaries. Values without outliers are in parenthesis.}
\begin{tabular}{@{}llllllll}
\br
Frequency & Datapoints & $a$ & $b$ & $R^2$ \\
\mr
$f_{h, mrg}$         & $158(156)$ & $4.1937(4.2114)$ & $-0.1425(-0.1470)$ & $0.944(0.979)$ \\
$f_{\Psi_4, mrg}$ & $102(100)$ & $4.5427(4.5748)$ & $-0.1889(-0.1970)$ & $0.884(0.934)$ \\
$f_{h, 2}$             & $106(104)$ & $4.4594(4.5131)$ & $-0.1578(-0.1711)$ & $0.792(0.854)$ \\
$f_{\Psi_4, 2}$     & $56(54)$ & $4.5440(4.6694)$ & $-0.1810(-0.2126)$ & $0.791(0.930)$ \\
$f_{h, max}$        & $154$ & $5.6331$ & $-0.3909$ & $0.252$ \\
$f_{\Psi_4, max}$ & $90$ & $4.9066$ & $-0.2357$ & $0.569$  \\
$\Delta t_{h}$ & $154(151)$ & $-1.1253(-1.1557)$ & $0.1368(0.1443)$ & $0.852(0.911)$ \\
$\Delta t_{\Psi_4}$ & $93$ & $-1.7257$ & $0.2324$ & $0.577$ \\ 
\br
\end{tabular}
\end{table}

\begin{table}[h]
\centering
\caption{\label{tab:tabA2}Summary of fit parameters to QUR for key frequencies and transient times across all codes, including unequal-mass binaries.}
\begin{tabular}{@{}lllllll}
\br
Frequency & Datapoints & $a$ & $b_0$ & $b_1$ & $b_2$ & $R^2$ \\
\mr
$f_{h, mrg}$         & $271$ & $4.1045$ & $-0.1797$ & $0.0596$ & $0.0019$ & $0.753$ \\
$f_{\Psi_4, mrg}$ & $209$ & $4.3507$ & $-0.3561$ & $0.3906$ & $-0.1700$ & $0.758$ \\
$f_{h, 2}$             & $193$ & $4.3879$ & $-0.0928$ & $-0.1240$ & $0.0774$ & $0.709$ \\
$f_{\Psi_4, 2}$     & $137$ & $4.3482$ & $-0.0800$ & $-0.1172$ & $0.0677$ & $0.544$ \\
$f_{h, max}$        & $267$ & $5.1602$ & $-1.0574$ & $1.7435$ & $-0.9475$ & $0.131$ \\
$f_{\Psi_4, max}$& $197$ & $4.6976$ & $-0.6929$ & $1.1658$ & $-0.6578$ & $0.144$\\
$t_{h,f_{max}}$ & $267$ & $-1.0614$ & $0.3608$ & $-0.5383$ & $0.2973$ & $0.669$ \\
$t_{\Psi_4,f_{max}}$ & $196$ & $-1.5297$ & $0.3228$ & $-0.2950$ & $0.1499$ & $0.476$ \\
\br
\end{tabular}
\end{table}

\begin{table}[h]
\centering
\caption{\label{tab:tabA3}QUR fit parameters for the strain $f_{h, mrg}$, with individual codes. Values without outliers are shown in parenthesis, and $\sum b_{q=1}$ obtained for equal-mass binaries.}
\begin{tabular}{@{}llllllll}
\br
Code(s) & Datapoints & $a$ & $b_0$ & $b_1$ & $b_2$ & $\sum b_{q=1}$ & $R^2$ \\
\mr
\texttt{SACRA} & $46$ & $4.2152$ & $-0.2852$ & $0.2394$ & $-0.1002$ & $-0.1460$ & $0.999$ \\
\ $\%$ Error\  & - & $2.70\%(0.09\%)$ & -- & -- & -- & $(0.68\%)$ & -- \\
\texttt{SpEC} & $6$ & $4.1825$ & $-0.4378$ & $0.5864$ & $-0.2847$ & $-0.1361$ & $0.976$ \\
\ $\%$ Error\  & - & $1.90\%(-0.69\%)$ & -- & -- & -- & $(7.41\%)$ & -- \\
\texttt{Whisky} & $57$ & $4.2262$ & $-0.5086$ & $0.7135$ & $-0.3552$ & $-0.1503$ & $0.982$ \\
\ $\%$ Error\  & - & $2.96\%(0.35\%)$ & -- & -- & -- & $(-2.24\%)$ & -- \\
\texttt{THC} & $97$ & $4.0572$ & $-0.2482$ & $0.2086$ & $-0.0632$ & $-0.1028$ & $0.680$ \\
\ $\%$ Error\  & - & $-1.15\%(-3.66\%)$ & -- & -- & -- & $(30.07\%)$ & -- \\
\texttt{BAM} & $65$ & $4.1585$ & $-0.1157$ & $-0.0961$ & $0.0776$ & $-0.1342$ & $0.734$ \\
\ $\%$ Error\  & - & $1.31\%(-1.26\%)$ & -- & -- & -- & $(8.71\%)$ & -- \\
\texttt{No THC} & $174$ & $4.1885$ & $-0.1130$ & $-0.1286$ & $0.1007$ & $-0.1409$ & $0.843$ \\
\ $\%$ Error\  & - & $2.05\%(-0.54\%)$ & -- & -- & -- & $(4.15\%)$ & -- \\
 No \texttt{BAM}\&\texttt{THC} & $109$ & $4.2253$ & $-0.2935$ & $0.2550$ & $-0.1113$ & $-0.1498$ & $0.987$ \\
\ $\%$ Error\  & - & $2.94\%(0.33\%)$ & -- & -- & -- & $(-1.90\%)$ & -- \\
\br
\end{tabular}
\end{table}

\begin{table}[h]
\centering
\caption{\label{tab:tabA4}QUR fit parameters for $f_{\Psi_4, mrg}$, with individual codes. For equal-mass binaries, the $b_i$ terms are combined in $\sum b_{q=1}$}
\begin{tabular}{@{}llllllll}
\br
Code(s) & Datapoints & $a$ & $b_0$ & $b_1$ & $b_2$ & $\sum b_{q=1}$ & $R^2$ \\
\mr
\texttt{SACRA} & $46$ & $4.5107$ & $-0.5433$ & $0.7299$ & $-0.3689$ & $-0.1823$ & $0.976$ \\
\texttt{THC} & $95$ & $4.3943$ & $-0.9572$ & $1.6642$ & $-0.8514$ & $-0.1444$ & $0.556$ \\
\texttt{BAM} & $69$ & $4.3480$ & $-0.3882$ & $0.5177$ & $-0.2651$ & $-0.1356$ & $0.769$ \\
No \texttt{THC} & $121$ & $4.3739$ & $-0.3291$ & $0.3343$ & $-0.1487$ & $-0.1435$ & $0.792$ \\
\texttt{SACRA}\&\texttt{SpEC} & $52$ & $4.5102$ & $-0.4540$ & $0.5226$ & $-0.2511$ & $0.1825$ & $0.967$ \\
\br
\end{tabular}
\end{table}

\begin{table}[h]
\centering
\caption{\label{tab:tabA5}Fit parameters for the $f_{max}$ quasi-universal relation from $\Psi_4$ with mass ratio dependence for the individual codes.}
\begin{tabular}{@{}llllllll}
\br 
Code(s) & Datapoints & $a$ & $b_0$ & $b_1$ & $b_2$ & $\sum b_{q=1}$ & $R^2$ \\
\mr   
\texttt{SACRA} & $46$ & $4.8691$ & $0.807$ & $-0.7848$ & $0.4810$ & $-0.2231$ & $0.912$ \\
\texttt{THC} & $77$ & $5.0011$ & $-0.1995$ & $0.3370$ & $-0.2696$ & $-0.2669$ & $0.157$ \\
\texttt{BAM} & $68$ & $4.5021$ & $-1.0646$ & $2.2822$ & $-1.3554$ & $-0.1367$ & $0.141$ \\
\texttt{SACRA}\&\texttt{SpEC} & $52$ & $4.8703$ & $0.0702$ & $-0.7620$ & $0.4687$ & $-0.2231$ & $0.914$ \\
\br
\end{tabular}
\end{table}

\begin{table}[h]
\centering
\caption{\label{tab:tabA6}Fit parameters for the $f_2$ quasi-universal relation from strain with mass-ratio dependence, for individual codes. }
\begin{tabular}{@{}llllllll}
\br        
Code(s) & $n$ & $a_0$ & $b_0$ & $b_1$ & $b_2$ & $\sum_n b_n$ & $R^2$ \\
\mr
\texttt{SACRA} & $40$ & $4.6033$ & $-0.1393$ & $-0.1515$ & $0.0935$ & $-0.1973$ & $0.968$ \\
            \ $\%$ Error\  & - & $4.91\%(2.00\%)$ & $-50.11\%$ & $-22.18\%$ & $20.80\%$ & $-15.31\%$ & - \\
\texttt{Whisky} & $50$ & $4.3761$ & $-0.6635$ & $1.1249$ & $-0.5977$ & $-0.1363$ & $0.827$ \\
            \ $\%$ Error\  & - & $-0.27\%(-3.04\%)$ & $-615.0\%$ & $1007\%$ & $-872.2\%$ & $20.34\%$ & - \\
\texttt{THC} & $62$ & $4.2107$ & $0.0996$ & $-0.4568$ & $0.2688$ & $-0.0884$ & $0.315$ \\
            \ $\%$ Error\  & - & $-4.04\%(-6.70\%)$ & $207.3\%$ & $-268.4\%$ & $247.3\%$ & $48.33\%$ & - \\
\texttt{BAM} & $41$ & $4.5002$ & $-0.2117$ & $0.1360$ & $-0.0952$ & $-0.1709$ & $0.863$ \\
            \ $\%$ Error\  & - & $2.56\%(-0.29\%)$ & $-128.1\%$ & $209.7\%$ & $-223.0\%$  & $0.12\%$ & - \\
No \texttt{THC} & $131$ & $4.4493$ & $-0.1186$ & $-0.0927$ & $0.0556$ & $-0.1557$ & $0.841$ \\
            \ $\%$ Error\  & - & $1.40\%,(-1.41\%)$ & $-27.80$ & $25.24\%$ & $-28.17\%$ & $9.00\%$ & - \\
\br
\end{tabular}
\end{table}

\begin{table}[h]
\centering
\caption{\label{tab:tabA7}Fit parameters for the $f_2$ quasi-universal relation fits from $\Psi_4$ with mass ratio dependence for the individual codes. \texttt{Whisky} and \texttt{SpEC} are excluded. }
\begin{tabular}{@{}llllllll}
\br 
Code(s) & $n$ & $a_0$ & $b_0$ & $b_1$ & $b_2$ & $\sum_n b_n$ & $R^2$ \\
\mr            
\texttt{SACRA} & $40$ & $4.5946$ & $-0.1634$ & $-0.892$ & $0.0576$ & $-0.195$ & $0.960$ \\ 
\texttt{THC} & $62$ &$4.2004$ & $0.1037$ & $-0.4562$ & $0.2666$ & $-0.0859$ & $0.304$ \\
\texttt{BAM} & $35$ & $4.4416$ & $-0.2503$ & $0.3014$ & $-0.2089$ & $-0.1578$ & $0.598$ \\
\br
\end{tabular}
\end{table}

\begin{table}[h]
\centering
\caption{\label{tab:tabA8}Fit parameters for the $t_{f_{max}}$ quasi-universal relation from strain, for the individual codes.}
\begin{tabular}{@{}llllllll}
\hline
Code(s) & $n$ & $a_0$ & $b_0$ & $b_1$ & $b_2$ & $\sum_n b_n$ & $R^2$ \\
\hline
\texttt{SACRA} & $46$ & $-1.2777$ & $0.2012$ & $-0.0442$ & $0.0160$ & $0.1730$ & $0.934$ \\
\texttt{Whisky} & $57$ & $-1.1663$ & $0.1156$ & $0.0939$ & $-0.0634$ & $0.1461$ & $0.951$ \\
\texttt{THC} & $93$ & $-1.0068$ & $0.4177$ & $-0.6741$ & $0.3602$ & $0.1038$ & $0.498$ \\
\texttt{BAM} & $65$ & $-1.0749$ & $0.2809$ & $-0.3533$ & $0.1987$ & $0.1263$ & $0.664$ \\
No \texttt{THC} & $174$ & $-1.1434$ & $0.3026$ & $-0.3679$ & $0.2068$ & $0.1415$ & $0.779$ \\
No \texttt{BAM}\&\texttt{THC} & $109$ & $-1.2098$ & $0.1907$ & $-0.0670$ & $0.0331$ & $0.1568$ & $0.936$ \\
\hline
\end{tabular}
\end{table}

\begin{table}[h]
\centering
\caption{\label{tab:tabA9}Fit parameters for the $t_{f_{max}}$ quasi-universal relation from $\Psi_4$, for the individual codes.}
\begin{tabular}{@{}llllllll}
\hline
Code(s) & $n$ & $a_0$ & $b_0$ & $b_1$ & $b_2$ & $\sum_n b_n$ & $R^2$ \\
\hline
\texttt{SACRA} & $46$ & $-1.4005$ & $0.4575$ & $-0.7035$ & $0.3969$ & $0.1509$ & $0.538$ \\
\texttt{THC} & $76$ & $-1.4852$ & $0.3782$ & $-0.4307$ & $0.2122$ & $0.1597$ & $0.220$ \\
\texttt{BAM} & $68$ & $-1.6111$ & $0.3549$ & $-0.3548$ & $0.2029$ & $0.2030$ & $0.598$ \\
No \texttt{THC} & $120$ & $-1.5473$ & $0.3084$ & $-0.2669$ & $0.1450$ & $0.1865$ & $0.573$ \\
\hline
\end{tabular}
\end{table}


\clearpage

\bibliographystyle{unsrt}
\bibliography{references}

\end{document}